\documentclass[12pt]{article}
\pdfoutput=1
\usepackage[DIV13]{typearea}
\usepackage{indentfirst}
\usepackage{amsmath}
\usepackage{amsfonts}
\usepackage{mathrsfs}
\usepackage{amssymb}
\usepackage{bbm}
\usepackage{epsfig}
\usepackage{graphicx}
\usepackage{slashed}
\usepackage{multicol}
\usepackage[usenames,dvipsnames]{color}
\usepackage{cite}
\RequirePackage[colorlinks=true,urlcolor=blue,anchorcolor=blue,citecolor=blue,filecolor=blue,
               linkcolor=blue,menucolor=blue,linktocpage=true,pdfproducer=medialab]{hyperref}
\usepackage{cancel}
\usepackage{feynmp}
\usepackage{array}
\usepackage{pdfpages}
\usepackage{fancyhdr}
\usepackage{ulem}
\newcommand{\email}[1]{\href{mailto:#1}{\tt #1}}

\numberwithin{equation}{section}
\newcommand{\LL}{\mathscr{L}}

\def\cL{{\cal L}}

\def\cF{{\cal F}}
\def\cO{{\cal O}}
\def\cP{{\cal P}}

\def\cY{{\bf Y}}

\def\be{\begin{equation}}
\def\ee{\end{equation}}
\def\beq{\begin{equation}}
\def\eeq{\end{equation}}
\def\bc{\begin{center}}
\def\ec{\end{center}}
\def\bea{\begin{eqnarray}}
\def\eea{\end{eqnarray}}

\def\nn{\nonumber}

\newcommand{\GeV}{\;\text{GeV}}

\renewcommand{\a}{\alpha}

\newcommand{\g}{\gamma}
\newcommand{\s}{\sigma}
\newcommand{\e}{\varepsilon}

\newcommand{\derp}{\partial}

\newcommand{\hc}{\mathrm{h.c.}}

\newcommand{\UH}{\mathbf{U}}

\newcommand{\VL}{\mathbf{V}}
\newcommand{\DL}{D}
\newcommand{\DLR}{\mathbf{D}}

\newcommand{\GF}{G_F}
\newcommand{\aem}{\alpha_\text{em}}

\newcommand{\tr}{{\rm Tr}}

\renewcommand{\to}{\rightarrow}

\newcommand{\WWd}{W_{\mu\nu}}
\newcommand{\ZZu}{Z^{\mu\nu}}
\newcommand{\ZZd}{Z_{\mu\nu}}

\newcommand{\st}{s_\theta}
\newcommand{\ct}{c_\theta}
\newcommand{\cdt}{c_{2\theta}}
\newcommand{\sdt}{s_{2\theta}}
\newcommand{\bmat}{\begin{pmatrix}}
\newcommand{\emat}{\end{pmatrix}}

\newcommand{\blue}[1]{\color{blue} #1 \color{black}}

\begin{document}
\begin{titlepage}
\vspace*{-1cm}
\phantom{hep-ph/***} 
{\flushleft
{\blue{FTUAM-14-17}}
\hfill{\blue{IFT-UAM/CSIC-14-039}}\\
{\blue{YITP-SB-14-12}}
\hfill{\blue{DFPD-2014/TH/11}}\\}
\vskip 1cm
\begin{center}
\mathversion{bold}
{\LARGE\bf Higgs ultraviolet softening}\\
\mathversion{normal}
\vskip .3cm
\end{center}
\vskip 0.5  cm
\begin{center}
{\large I.~Brivio}~$^{a)}$,
{\large O.\ J.\ P.\ \'Eboli}~$^{c)}$,
{\large M.B.~Gavela}~$^{a)}$,\\[2mm]
{\large M.\ C.\ Gonzalez--Garc\'ia}~$^{e,d,b)}$,
{\large L.~Merlo}~$^{a)}$,
{\large S.~Rigolin}~$^{f)}$
\\
\vskip .7cm
{\footnotesize
$^{a)}$~
Departamento de F\'isica Te\'orica and Instituto de F\'{\i}sica Te\'orica, IFT-UAM/CSIC,\\
Universidad Aut\'onoma de Madrid, Cantoblanco, 28049, Madrid, Spain\\
\vskip .1cm
$^{b)}$~
C.N.~Yang Institute for Theoretical Physics and Department of Physics and Astronomy, SUNY at Stony Brook, Stony Brook, NY 11794-3840, USA\\
\vskip .1cm
$^{c)}$~
Instituto de F\'{\i}sica, Universidade de S\~ao Paulo, S\~ao Paulo -- SP, Brazil\\
\vskip .1cm
$^{d)}$~
Departament d'Estructura i Constituents de la Mat\`eria and ICC-UB, Universitat de Barcelona, 647 Diagonal, E-08028 Barcelona, Spain\\
\vskip .1cm
$^{e)}$~
Instituci\'o Catalana de Recerca i Estudis Avan\c{c}ats (ICREA)\\
\vskip .1cm
$^{f)}$~
Dipartimento di Fisica e Astronomia ``G.~Galilei'', Universit\`a di Padova and \\
INFN, Sezione di Padova, Via Marzolo~8, I-35131 Padua, Italy
\vskip .3cm
\begin{minipage}[l]{.9\textwidth}
\begin{center} 
\textit{E-mail:} 
\email{ilaria.brivio@uam.es},
\email{eboli@fma.if.usp.br},
\email{belen.gavela@uam.es},
\email{concha@insti.physics.sunysb.edu},
\email{luca.merlo@uam.es},
\email{stefano.rigolin@pd.infn.it}
\end{center}
\end{minipage}
}
\end{center}
\vskip 0.5cm
\begin{abstract}
We analyze the leading effective operators which induce a quartic momentum dependence in the Higgs propagator, for a linear and for a non-linear realization of electroweak symmetry breaking. Their  specific study is relevant for the understanding of  the ultraviolet sensitivity to new physics.  
Two methods  of analysis are applied, trading the Lagrangian coupling by: i) a ``ghost" scalar, after the Lee-Wick procedure; ii) other effective operators via  the equations of motion.  The two paths are shown to lead to the same effective Lagrangian at first order in the operator coefficients.  It follows a modification of the Higgs potential and of the fermionic couplings in the linear realization, while in the non-linear one  anomalous  quartic gauge couplings, Higgs-gauge couplings and gauge-fermion interactions are  induced in addition. 
Finally, all LHC Higgs and other data presently available  are used to constrain the operator coefficients; the future impact of $pp\to\text{4 leptons}$ data via off-shell Higgs exchange and of vector boson fusion data is considered as well. 
For completeness, a summary of pure-gauge and gauge-Higgs signals exclusive to non-linear dynamics at leading-order is included.
 
\end{abstract}
\end{titlepage}
\setcounter{footnote}{0}

\tableofcontents

%
%

\newpage

\section{Introduction}

A revival of interest in theories with higher derivative kinetic terms~\cite{Lee:1970iw,Lee:1969fy} is taking place, as the increased momentum dependence of propagators 
softens the sensitivity to ultraviolet scales. 
Quadratic divergences are absent due to the faster fall-off of the momentum dependence of the propagators. For instance this avenue has been recently explored in view of an alternative solution to the electroweak hierarchy problem~\cite{Grinstein:2007mp,Espinosa:2011js}.

 Originally proposed by Lee and Wick~\cite{Lee:1970iw,Lee:1969fy}, a large literature followed to  ascertain the field theoretical consistency of this type of theories, in particular from the point of view of unitarity and causality. The issue is delicate as a second pole appears in the field propagators, and this pole has a wrong-sign residue. Naively such theories are unstable and not unitary. 
 The present understanding is that the S matrix for asymptotically free states may remain unitary, though, and acausality only occurs at the microscopic level while macroscopically and/or in any measurable quantity causality holds as it should.
 
 For the computation of physical amplitudes, a modification of the usual rules to compute perturbative amplitudes was proposed~\cite{Cutkosky:1969fq,Coleman:1969xz,Lee:1971ix,Nakanishi:1971jj} respecting the aforementioned desired properties. A  more user-friendly field-theory tool~\cite{Grinstein:2007mp} to approach these theories consists in trading the higher derivative kinetic term  by the presence of a new state with the same quantum numbers  of the standard field and quadratic kinetic energy, albeit with a ``wrong" sign for both quadratic terms (kinetic energy and mass), i.e. a state of negative norm: a Lee-Wick (LW) partner or ``ghost".  It corresponds to the second pole in the propagator, describing an unstable state that would thus not threaten the unitarity of the S matrix, as only the asymptotically free states participating in a scattering process are relevant for the latter.
 
 In this paper, we focus on the study of a higher derivative kinetic term for the Higgs particle, in a model independent way.  Although present Higgs data are fully consistent with the Higgs particle being part of a gauge SU(2) scalar doublet, the issue is widely open and all efforts should be done to settle it. Two main classes of effective Lagrangians
are pertinent, depending on how the Standard Model (SM) electroweak  symmetry breaking (EWSB)
is assumed to be realized in the presence of a light Higgs particle: linearly for an elementary Higgs particle~\cite{Buchmuller:1985jz,Grzadkowski:2010es,Hagiwara:1993ck} or
non-linearly for  a ``dynamical" -composite- light one~\cite{Bagger:1993zf,Koulovassilopoulos:1993pw,Burgess:1999ha,Grinstein:2007iv,Azatov:2012bz,Alonso:2012px,Alonso:2012pz,Buchalla:2013rka}. The relevant couplings to be added to the SM Lagrangian will be denoted by
\begin{equation}
\cO_{\square \Phi} = 
 (D_\mu D^\mu \Phi)^\dag \, (D_\nu D^\nu \Phi)
\label{OboxPhi}
 \end{equation}
for linearly realized electroweak symmetry breaking (EWSB) scenarios, and
\begin{equation}
\cP_{\square h}= \dfrac{1}{2}\square h \square h= \dfrac{1}{2}\,(\partial_\mu \partial^\mu h) \, (\partial_\nu \partial^\nu h)
\label{Oboxh}
\end{equation}
 if the light Higgs stems from non-linearly realized EWSB. In  Eq.~(\ref{OboxPhi}) $\Phi$ denotes the gauge $SU(2)$ scalar doublet, which in the unitary gauge reads  $\Phi=\left(0, (v+h)/\sqrt{2}\right)$ with $v/\sqrt2$ being the  $\Phi$  vacuum expectation value (vev) and $h$ the Higgs excitation. $D_\mu$ stands for the covariant derivative
 \begin{equation}
 D_\mu \Phi\equiv  \left( \partial_\mu + i g W_\mu + \dfrac{i\,g'}{2} B_\mu \right)\Phi
 \end{equation}
 with $W_\mu\equiv W_{\mu}^a(x)\sigma_a/2$ and $B_\mu$ denoting the
$SU(2)_L$ and $U(1)_Y$ gauge bosons, respectively.

In equation (\ref{Oboxh}),  $h$ denotes instead  a generic scalar singlet, whose couplings are described by a non-linear Lagrangian (often dubbed chiral Lagrangian) and do not need to match those of a $SU(2)$ doublet component. 

Note that the operators $\cO_{\square \Phi}$ and $\cP_{\square \Phi}$ are but rarely~\cite{Grzadkowski:2010es} considered by practitioners of effective Lagrangian analyses, and almost never selected as one of the elements of the operator bases. They tend to be substituted instead  by (a combination of) other operators --which include fermionic ones-- because the bounds on exotic fermionic couplings are often more stringent in constraining BSM theories than those from bosonic interactions. Nevertheless, the new data and the special and profound theoretical impact of higher derivative kinetic terms deserve focalised studies, to which this paper intends to contribute.

In this context it is important to notice that, in order to have any impact on the hierarchy problem, the validity of the operators under study should be extrapolated into the regime $E\gg \Lambda$, which is beyond the usual regime where EFT description is valid. In this sense, the SM Lagrangian with the addition of these operators can be treated as the complete Lagrangian in the ultraviolet.

Either in the linear or the non-linear realizations,  the contribution to the Lagrangian of the effective operators in Eqs.~(\ref{OboxPhi}) and (\ref{Oboxh}) can be parametrised as 
 \begin{equation}
 \delta \cL= c_i \cO_i \,,
 \label{deltaL}
 \end{equation}
 with $\cO_i\equiv\{\cO_{\square \Phi},\cP_{\square h}\}$ respectively, with the parameters $c_i$ having mass dimension $-2$.~\footnote{From the point of view of the chiral expansion, $\cP_{\square \Phi}$ is a four-derivative coupling, and a slightly different normalization (by a  $v^2$ factor) was adopted in Ref.~\cite{Brivio:2013pma}, using a dimensionless coefficient; the choice here allows to use the same notation for both expansions.\label{fotenote1}} The impact of $\cO_{\square \Phi}$ and $\cP_{\square h}$ appears as a correction in the propagator of the $h$ scalar which is quartic in four-momentum:
\begin{equation}
 \dfrac{i}{p^2-m_h^2+c_i\,p^4}\,.
 \label{quarticprop}
\end{equation}
This propagator has now two poles and describes thus two degrees of freedom. For instance for $1/c_i\gg m_h^2$ they are approximately located at~\cite{Grinstein:2007mp}
\beq
p^2=m_h^2 \hspace{3cm}\text{and}\hspace{3cm} p^2=-1/c_i\,,
\eeq
which implies that the sign of the operator coefficient needs  to obey $c_i<0$ in order to avoid tachyonic instabilities. 

 It is  important to
find signals which discriminate among those two categories --linear versus non-linear EWSB-- and this
will be one of the main focuses of this paper for the higher derivative scalar kinetic terms considered. It will be shown that the effects of the couplings in Eqs. (\ref{OboxPhi}) and  (\ref{Oboxh})  differ on their implications for the gauge and gauge-Higgs sectors. The phenomenological analysis will be restricted to tree-level effects and consistently to first order in $c_i$, and we will use two independent and alternative techniques, showing that they lead to the same results:  

\begin{itemize}
\item[-] To trade the higher-derivative coupling by a LW ``ghost" heavy particle, which is subsequently integrated out.
\item[-]  To apply first the Lagrangian equations of motion (EOM) to the operator, trading the coupling by other standard higher-dimension effective operators, which only require traditional fields and field-theory methods.
\end{itemize}
Together with exploring the different  physical effects expected from the Higgs linear higher-derivative term $\cO_{\square \Phi}$ and  the non-linear one $\cP_{\square h}$, we will clarify their exact theoretical relation, determining which specific combination of non-linear operators would result in the same physics impact than the linear operator $\cO_{\square \Phi}$.

The phenomenological analysis below includes  as well a study of the impact of both operators in present and future LHC data. In the case of the LW version of the SM, it has been shown~\cite{Alvarez:2008za}  that the measurements of the S and T parameters set very strong constraints on the gauge and fermionic LW partner masses, which  need to exceed several TeV; this implies a sizeable tension with the issue of the electroweak hierarchy problem, as the LW partners induce a finite shift in the Higgs mass proportional to their own masses.  On the contrary, the EW constraints are mild for the  Higgs doublet LW partners, whose impact may be within LHC reach~\cite{Carone:2014kla}. We explore the experimental prospects for $\cO_{\square \Phi}$ and $\cP_{\square h}$ at first order in the effective operator coefficients,  focusing only on the quark sector for simplicity as the extension to the lepton sector is straightforward.

The structure of the manuscript can be easily inferred from the Table of Contents.
 
%
%
 \boldmath
 \section{Elementary Higgs: $\cO_{\square \Phi}$}
\unboldmath

The  quark-Higgs sector of the SM Lagrangian supplemented by $\cO_{\square \Phi}$ will be considered in this section:
\begin{equation}
\mathcal{L} = (D_\mu \Phi)^\dag D^\mu \Phi
-\left(\bar{q}_L\tilde{\Phi}Y_U u_R+\bar{q}_L\Phi Y_D d_R+\text{h.c.}\right)+\,c_{\square \Phi}\,\cO_{\square \Phi} - V(\Phi^\dag\Phi)\,,
\label{Lagl}
\end{equation}
where $\tilde{\Phi}\equiv i \sigma_2 \Phi$, and the Standard Model potential,
\begin{equation}\label{linear_potential}
 V(\Phi^\dag\Phi)= \lambda\,  \left[\Phi^\dag\Phi\,- \frac{v^2}{2}\right]^ 2\,,
\end{equation}
can be rewritten for future convenience in the unitary gauge in terms of the Higgs particle mass, $m_h^2= 2\lambda v^2$ and the Higgs doublet vev $\langle\Phi\rangle= v/\sqrt{2}$  as
\begin{equation}
 V(h) =\dfrac{m_h^2}{2}h^2+\dfrac{m_h^2}{2v}h^3+\dfrac{m_h^2}{8v^2}h^4 \,.
 \label{VSM}
\end{equation}

\subsection{Analysis in terms of the LW  ghost}
\label{Linearghost}
   The Lee-Wick method for the case of a complex scalar doublet is applied next to the analysis of the operator $\cO_{\square \Phi}$ in Eqs.~(\ref{OboxPhi}) and (\ref{deltaL}), following Ref.~\cite{Grinstein:2007mp}. Defining an auxiliary complex $SU(2)$ doublet $\varphi$, Eq.~(\ref{Lagl}) can be rewritten as a two-scalar-field Lagrangian: 
\begin{equation}
\begin{aligned}
 \mathcal{L} =& (D_\mu \Phi)^\dag D^\mu \Phi+(D_\mu \varphi)^\dag D^\mu \Phi+(D_\mu \Phi)^\dag D^\mu \varphi+\\
 &-\left(\bar{q}_L\tilde{\Phi}Y_U u_R+\bar{q}_L\Phi Y_D d_R+\text{h.c.}\right)-\dfrac{1}{c_{\square\Phi}}\varphi^\dag\varphi-V(\Phi^\dag\Phi)\,.
 \end{aligned}
 \label{LBoxPhiLW}
\end{equation}
The mass squared term for the auxiliary field is given by $-1/{c_{\square\Phi}}$, which requires $c_{\square\Phi}<0$ to avoid a tachyonic resonance. The kinetic energy terms can now be diagonalised via  the simple field redefinitions  $\Phi\to\Phi'-\varphi'$, $\varphi\to \varphi'$, and the mass terms can be  diagonalised by a subsequent symplectic rotation given by:
\begin{equation}
  \bmat \Phi'\\ \varphi'\emat = \bmat \cosh\alpha& \sinh\alpha\\ \sinh\alpha& \cosh\alpha \emat \bmat \Phi''\\ \varphi''\emat \,,
  \label{rotation}
\end{equation}
where
\begin{equation}
\tanh2\alpha = \dfrac{2x}{1+2x}\,,\qquad  \text{with\,} \quad x\equiv - c_{\square\Phi} \,m_h^2/2\,.
\label{mixing_linear}
\end{equation}
 Finally, dropping the primes on the field notation, the scalar Lagrangian in Eq.~(\ref{LBoxPhiLW}) can be rewritten as   \begin{align}
 \mathcal{L}^{\varphi, \Phi} =&(D_\mu\Phi)^\dag D^\mu\Phi-(D_\mu\varphi)^\dag D^\mu\varphi
 +\mathcal{L}^\varphi_Y-V(\Phi,\varphi)
 \label{LvarphiPhi}
\end{align}
with
\begin{align}
 \mathcal{L}^\varphi_Y&=-\left(1+x\right) \left(\bar{q}_L(\tilde{\Phi}-\tilde{\varphi})Y_U u_R+\bar{q}_L(\Phi-\varphi) Y_D d_R+\text{h.c.}\right)\,,\\
 V(\Phi, \varphi)&=-\frac{m_h^2}{2}\left(1-x+\frac{1}{x}\right)\varphi^\dag\varphi-\frac{m_h^2}{2}\left(1-x\right)\Phi^\dag\Phi
+\dfrac{m_h^2}{2v^2}\left(1-4x\right)\left((\Phi-\varphi)^\dag(\Phi-\varphi)\right)^2\,,
 \label{VPhiphi}
\end{align}
expanded at order $x$, assuming small $x$ values. 
 The location of the minimum of the Higgs potential gets $c_{\square\Phi}$ corrections. For instance, for a BSM scale large compared with the Higgs mass (i.e. $x\to 0$), the approximate location of the vacuum   corresponds to:
\begin{align}
\Phi &\rightarrow  \langle\Phi\rangle+ \frac{h}{\sqrt2}\,,&  
\langle\Phi\rangle&=\frac{v}{\sqrt{2}} \left(1 + \frac{15}{2}x^2\right) +\mathcal{O}(x^3)\,, \label{VminimPhi}\\[3mm]
 \varphi &\rightarrow \langle\varphi\rangle+ \frac{\chi}{\sqrt2}\,,&
\langle\varphi\rangle&=-x\,\frac{v}{\sqrt{2}} (1 - 2x) +\mathcal{O}(x^3)\,,
  \label{Vminimphi}
\end{align}
 where $h$ and $\chi$ are the  field excitation over the potential minima, and the exact potential has been retaken and terms up to $x^2$ considered. 
 In consequence, at leading order in $c_{\square\Phi}$ the minimum of the Higgs potential remains unchanged. 
For the sake of comparison with the non-linear case in the next section, it is useful to write explicitly the potential restricted to the  $h$ and    
 $\chi$ fields.
 After a further necessary diagonalization of the  $h$ and $\chi$ dependence, their scalar potential reads at first order in $x$:
 \beq
 \begin{split}
V(h,\chi) &= \frac{m_h^2}{2}(1+2x )h^2+
\frac{m_h^2}{2}\left(1+2x-\frac{1}{2x}\right)\chi^2
+\frac{m_h^2}{2v}(1+6x)(h-\chi)^3+\\
&\quad
+\frac{m_h^2}{8v^2}(1+8x)(h-\chi)^4\,.
\end{split}
\label{Vlinear_hchi}
\eeq
Eqs.~(\ref{LvarphiPhi}) and (\ref{Vlinear_hchi}) illustrate that for small $x$ the $\chi$ state exhibits a ``wrong" sign in both the kinetic energy and the mass terms.

\subsubsection*{Integrating out the heavy scalar}

At first order in the operator coefficient $c_{\square\Phi}$, the mixing in Eq.~(\ref{mixing_linear}) may  be approximated by  $\tanh\alpha \sim2x=-c_{\square\Phi}\,m_h^2$, and 
the effect of the negative-norm heavy field described by $\varphi$  with absolute mass$~\sim |c_{\square\Phi}^{-1}|$ can be integrated out via its EOM:
\beq
 \bar{\varphi_i} =c_{\square\Phi}\left(\bar{d}_R Y^\dag_D q_{L,i}+\bar{q}_{L,j} \e_{ji} Y_U u_R +\dfrac{m_h^2}{v^2}(\Phi^\dag\Phi)\Phi_i\right)+\mathcal{O}\left(c_{\square\Phi}^2\right)\,,
\label{EOMvarphi}
\eeq
Throughout the paper we will work on the so-called Z-scheme of renormalization, in which the 
five relevant electroweak parameters of the SM Lagrangian (neglecting fermion masses), $g_s$, $g$,
$g'$, $v$ and the $h$ self-coupling, are fixed from the following five  observables: the world average value of $\a_s$~\cite{Beringer:1900zz}, 
 the Fermi constant $G_F$ as extracted from muon decay~\cite{Beringer:1900zz}, 
$\aem$ extracted from Thomson scattering~\cite{Beringer:1900zz}, 
$m_Z$ as determined from the $Z$ lineshape at LEP I
~\cite{Beringer:1900zz}, and 
$m_h$ from the present LHC measurement~\cite{Aad:2013wqa,Chatrchyan:2013lba}. Eq.~(\ref{EOMvarphi}) above indicates that $O_{\square\Phi}$ will impact the renormalised fermion masses and the Higgs sector parameters. Specifically for the latter, while the electroweak vev $v \equiv (\sqrt{2}\GF)^{-1/2}$ is not corrected, 
the Higgs mass renormalization must absorb a correction
\begin{equation}
 \frac{\delta m_h^2}{m_h^2} = 2x\,.
\end{equation}
The resulting  renormalized effective Lagrangian reads (omitting again fermionic and gauge kinetic terms):
\begin{equation}\label{L_renorm_linear_fermions}
\begin{aligned}
 \mathcal{L}_{\square\Phi} &=\left(D_\mu\Phi\right)^\dag D^\mu \Phi+ \mathcal{L}_{\square\Phi}^Y+ \mathcal{L}_{\square\Phi}^{4F}- V_{\square\Phi}\,,
\end{aligned}
\end{equation}
where
\begin{align}
 \begin{split}
 \mathcal{L}_{\square\Phi}^Y=&-\left[\bar{q}_L\tilde{\Phi}Y_U u_R+\bar{q}_L\Phi Y_D d_R+\text{h.c.}\right]\left(1-x\left(1-2\frac{\Phi^\dag\Phi}{v^2}\right)\right)\underrightarrow{\text{\scriptsize unitary gauge}}\\
 &-\dfrac{v+h}{\sqrt2}\left[\bar{u}_LY_U u_R+\bar{d}_LY_D d_R+\text{h.c.}\right]\left(1+\dfrac{x}{v^2}(h^2+2hv)\right)\,,\label{LsquarePhi}
 \end{split}\\[5mm]
 \begin{split}
 \mathcal{L}_{\square\Phi}^{4F}=&-x\dfrac{2}{m_h^2}\Big[
 +(\bar{u}_R Y^\dag_U d_L)(\bar{d_L} Y_U u_R)+(\bar{u}_R Y^\dag_U u_L)(\bar{u}_LY_U u_R)\\
 &+(\bar{u}_L Y_D d_R)(\bar{d}_R Y^\dag_D u_L) 
 +(\bar{d}_L Y_D d_R)(\bar{d}_R Y^\dag_D d_L)\\
 &+\Big\{(\bar{u}_L Y_U u_R)(\bar{d}_L Y_D d_R)
 -(\bar{d}_L Y_U u_R)(\bar{u}_L Y_D d_R)
 + \text{h.c.}\Big\}\Big]\,,
  \end{split}\\[5mm]
  \begin{split}
 V_{\square\Phi}=&-\dfrac{m_h^2}{2}\left(1-3x\right)\Phi^\dag \Phi
 +\dfrac{m_h^2}{2v^2}\left(1-6x \right)\left(\Phi^\dag \Phi\right)^2
 +2x\dfrac{  m_h^2}{v^4 }\left(\Phi^\dag \Phi\right)^3\,\underrightarrow{\text{\scriptsize unitary gauge}}\\
 &\dfrac{m_h^2}{2}h^2+\dfrac{m_h^2}{2v}\left(1+4x\right)h^3
 +\dfrac{m_h^2}{8v^2}\left(1+24 x\right)h^4+ x \dfrac {m_h^2}{2v^3}\left( 3 h^5+\dfrac{ 1}{2v }h^6\right)\,.
\label{VboxPhi}
 \end{split}
\end{align}
 It follow deviations from SM expectations in fermion-Higgs couplings, four-fermion interactions and scalar properties; in particular, the relation between the Higgs self-couplings and its mass is different from the SM one; this fact can be directly probed at the LHC and ILC~\cite{Efrati:2014uta}. Moreover, the Higgs potential exhibits now $h^5$ and $h^6$ terms not present in the SM,  which require $c_{\square\Phi}<0$ for stability, consistently with the arguments given in the Introduction. Note as well that,  {\it for the linear realization of EWSB under discussion, the couplings involving gauge particles are not modified with respect to their SM values}.

\subsection{Analysis via EOM}

An avenue alternative to the LW method when working at first order in the operator coefficient, and one which involves  only standard fields and standard field theory rules,  is to apply directly the EOM for the $\Phi$ field to the operator $\cO_{\square \Phi}$ in Eq.~(\ref{Lagl}):
\begin{align}\label{EOM_phi}
 \square \Phi_i &= -\dfrac{\delta V}{\delta(\Phi^\dag\Phi)} \Phi_i-\left(\bar{d}_R Y^\dag_D q_{L,i}+\bar{q}_{L,j} \e_{ji} Y_U u_R\right)\,,\\
 \square \Phi^\dag_i &= -\Phi^\dag_i\dfrac{\delta V}{\delta(\Phi^\dagger\Phi)} -\left(-\bar{u}_R Y^\dag_U \e_{ij} q_{L,j}+\bar{q}_{L,i} Y_D d_R\right)\,. 
\end{align}
We have checked that this method leads to the same low-energy renormalized effective Lagrangian than that in Eqs.~\eqref{L_renorm_linear_fermions}-\eqref{VboxPhi}, obtained via the Lee-Wick procedure involving a ``ghost" field.

\subsubsection*{Higgs potential}

\begin{figure}[h!]
\centering
 \includegraphics[width=.8\textwidth]{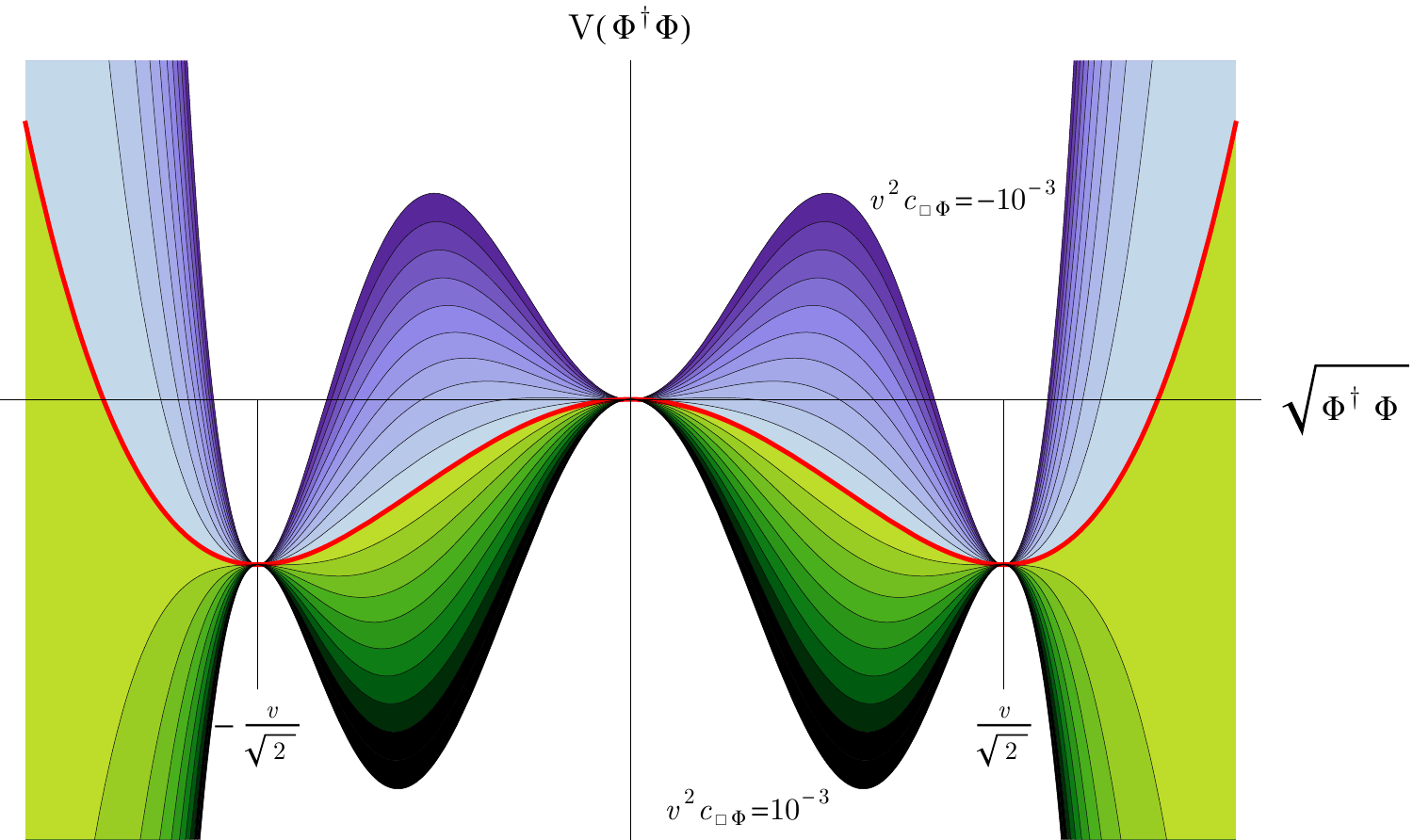}
 \caption{\it The scalar potential in the linear Lagrangian for different values of the coefficient $c_{\square \Phi} v^2$. The solid red line denotes the SM and the interline spacing is $\Delta(v^2 c_{\square\Phi})=7.5\cdot 10^{-5}$.}\label{fig.linearV}
\end{figure}

Figure~\ref{fig.linearV} shows the dependence of the scalar potential on $c_{\square \Phi}$: the points $|\Phi|=\pm v/\sqrt{2}$, corresponding to the SM vacuum, switch from stable minima to maxima as $c_{\square \Phi}$ runs from negative to positive values. The location of Higgs vev for negative $c_{\square \Phi}$ is not modified at this order, see Eq.~(\ref{VminimPhi}).

%
%

\boldmath
\section{Light dynamical Higgs: $\cP_{\square h}$}
\label{Composite_section}
\unboldmath

This section deals with the alternative scenario of a light dynamical Higgs, whose CP-even bosonic effective Lagrangian has been discussed in Refs.~\cite{Alonso:2012px,Brivio:2013pma}. For simplicity and focus, the leading-order Lagrangian  will be taken to be that of the SM modified only by  the action of the operator $\cP_{\square h}$ in Eq.~(\ref{Oboxh}).
The scalar potential will thus be assumed as well to take the SM form for $h$, to facilitate comparison with the linear case; nevertheless, in Appendix~\ref{Generic_chiral_potential} we discuss the extension to the case of a completely general potential for a singlet scalar field $h$, showing that the conclusions obtained below are maintained.

The quark-Higgs sector of the Lagrangian  reads then
\begin{align}
\mathcal{L} =& \frac{1}{2}\derp_\mu h\derp^\mu h-\dfrac{(v+h)^2}{4}\tr[\VL_\mu\VL^\mu]-\dfrac{v+h}{\sqrt2}(\bar{Q}_L \UH\cY Q_R+\text{h.c.})+c_{\square h}\cP_{\square h}- V(h)\,,
\label{Lchiral}
\end{align}
where $V(h)$  takes the functional form in Eq.~(\ref{VSM}).  
 \mbox{$\VL_\mu\equiv \left(\DLR_\mu\UH\right)\UH^\dagger$}, where $\UH(x)$ is a dimensionless unitary matrix describing the longitudinal degrees of freedom of the EW
gauge bosons:
\beq
\UH(x)=e^{i\sigma_a \pi^a(x)/v}\, , \qquad \qquad  \UH(x) \rightarrow L\, \UH(x) R^\dagger\,,
\eeq
where $L$, $R$
denotes $SU(2)_{L,R}$ global transformations, respectively. $\VL_\mu$ is thus a vector chiral field belonging to the adjoint of the global  $SU(2)_L$ symmetry, and 
the covariant derivative reads 
\beq
\DLR_\mu \UH(x) \equiv \derp_\mu \UH(x) +igW_{\mu}(x)\UH(x) - 
                      \dfrac{ig'}{2} B_\mu(x) \UH(x)\sigma_3 \, .
\eeq
Note that Eq.~(\ref{Lchiral}) is simply the SM Lagrangian written in chiral notation, but for the additional presence of the $\cP_{\square h}$ coupling.

\subsection{Analysis in terms of the LW ghost}

In parallel to the analysis in Sect.~(\ref{Linearghost}), for $c_{\square h}<0\,$ the action of the operator $\cP_{\square h}$ in the Lagrangian Eq.~(\ref{Lchiral}) can be traded for that of an auxiliary SM singlet scalar field $\chi$, and the Lagrangian in Eq.~(\ref{Lchiral}) reads then 
\begin{equation}
\mathcal{L} = \frac{1}{2}\derp_\mu h\derp^\mu h+\derp_\mu h\derp^\mu\chi -\frac{(v+h)^2}{4}\tr[\VL_\mu\VL^\mu] -\dfrac{v+h}{\sqrt2}(\bar{Q}_L \UH\cY Q_R+\text{h.c.})- V(h,\chi)\,,
\end{equation}
where the non-scalar kinetic terms were omitted and (see Appendix~\ref{Generic_chiral_potential})
\begin{equation}\label{V_chiral}
V(h,\chi)=  \frac{m_h^2}{2}h^2+\frac{m_h^2}{2v}h^3+\frac{m_h^2}{8v^2} h^4+\frac{1}{2c_{\square h}}\chi^2\,.
\end{equation}
The kinetic energy terms are diagonalised via  the  field redefinitions  $h\to h'-\chi'$, $\chi\to \chi'$, and the mass terms can be then diagonalised by a subsequent symplectic rotation 
analogous to that in  Eq.~(\ref{rotation}) (with $\Phi$ and $\varphi$ replaced by $h$ and $\chi$, respectively), with a mixing angle given by 
\begin{equation}
 \tanh2\alpha = \dfrac{-4x}{1-4x}\,,\qquad  \text{with\,} \quad x\equiv - c_{\square h} \,m_h^2/2\,.
\label{mixing_chiral}
\end{equation}
Finally, dropping the primes on the field notation and omitting again fermionic and gauge kinetic terms, the Lagrangian reads:
\begin{equation}
 \mathcal{L}^{h,\chi} = \dfrac{1}{2}\derp_\mu h\derp^\mu h-\dfrac{1}{2}\derp_\mu \chi\derp^\mu \chi
 +  \mathcal{L}^\chi_Y + \mathcal{L}^\chi_\text{gauge} -  V(h,\chi)\,,
\end{equation}
where, at first order in $x$,
\begin{align}
  \mathcal{L}^\chi_Y=&-\dfrac{1}{\sqrt2}(\bar{Q}_L \UH\cY Q_R+\text{h.c.})\left[v+ (h-\chi)\left(1+2x\right)\right]\,,\\[3mm]
  \mathcal{L}^\chi_\text{gauge} =&-\dfrac{1}{4}\tr[\VL_\mu\VL^\mu]\left[v^2+2v(1+2x)(h-\chi)+(1+4x)(h-\chi)^2\right]\,,
\end{align}
while the scalar potential $V(h,\chi)$ coincides with that given  in Eq.~(\ref{Vlinear_hchi}).

\subsubsection*{Integrating out the heavy scalar}

For small  $x$ (that is,  $\chi$ mass large compared to the Higgs mass), the first order EOM can be used  to integrate out the LW partner,
\begin{align}
 \bar{\chi} &= \dfrac{c_{\square h}}{2}\left[\sqrt2(\bar{Q}_L \UH\cY Q_R+\text{h.c.})+
 \tr[\VL_\mu\VL^\mu](v+h)+\dfrac{m_h^2}{v^2}h^2(h+3v)\right]+\mathcal{O}(c_{\square h}^2)\,.
\end{align}
While the masses of the gauge and fermion fields are unaffected by the presence of $\cP_{\square h}$,  the Higgs mass renormalization absorbs the correction
\begin{equation}
 \frac{\delta m_h^2}{m_h^2} =  2x \,.
\end{equation}
The resulting effective Lagrangian for the $h$ field is given by (omitting kinetic terms other than the Higgs one)
\begin{equation}\label{L_renorm_chiral_ferm}
 \begin{aligned}
 \mathcal{L}_{\square h} &=\dfrac{1}{2}\derp_\mu h\derp^\mu h+\mathcal{L}_{\square h}^Y+\mathcal{L}_{\square h}^{4F}+\mathcal{L}_{\square h}^\text{gauge}- V_{\square h}(h)\,,
\end{aligned}
\end{equation}
with
\begin{align}
 \mathcal{L}_{\square h}^Y=&-\dfrac{v+ h}{\sqrt2}\left(\bar{Q}_L\UH \cY Q_R+\text{h.c.}\right)\left(1+\dfrac{x}{v^2}(h^2+2vh)\right)+\nn\\
 & -\dfrac{x}{m_h^2}\dfrac{(v+h)}{\sqrt2 }\tr[\VL_\mu\VL^\mu] \left(\bar{Q}_L\UH\cY Q_R+\text{h.c.}\right)\,,\\[5mm]
  \mathcal{L}_{\square h}^{4F}=&- \dfrac{x}{2m_h^2}\left(\bar{Q}_L\UH\cY Q_R+\text{h.c.}\right)^2\,,\\[5mm]
 \mathcal{L}_{\square h}^\text{gauge}=&-\dfrac{(v+h)^2}{4}\tr[\VL_\mu\VL^\mu]\left(1+2\dfrac{x}{v^2}(h^2+2vh)\right)-\dfrac{x}{4m_h^2}\tr[\VL_\mu\VL^\mu]^2(v+h)^2\,,\\[5mm]
 V_{\square h}(h)=&\dfrac{m_h^2}{2}h^2+\dfrac{m_h^2}{2v}\left(1+4x\right)h^3
 +\dfrac{m_h^2}{8v^2}\left(1+24x\right)h^4+ x\dfrac{m_h^2}{2v^3}\left(3 h^5+ \dfrac{1}{2v}h^6\right)\,.
 \label{Vhchiral}
\end{align}
$\mathcal{L}_{\square h}^Y$ above shows that anomalous gauge-fermion interactions weighted by Yukawas are expected in the non-linear realization, in addition
 to the pure Yukawa-like corrections present  in the linear expansion, see Eq.~(\ref{LsquarePhi}). Furthermore,  the potential $V_{\square h}(h)$ in Eq.~(\ref{Vhchiral}) matches exactly the potential in Eq.~(\ref{VboxPhi}) for the linear case, as it should, exhibiting higher than quartic Higgs couplings that requires $c_{\square h}<0$ (i.e., $x>0$) for the stability of the potential.

 In summary, the resulting effective Lagrangian for the non-linear case in Eqs.~\eqref{L_renorm_chiral_ferm}-\eqref{Vhchiral} shows deviations from SM expectations in fermion-Higgs couplings, four-fermion interactions and scalar properties, a pattern already found in the previous section for an elementary Higgs. Nevertheless, important distinctive features appear with respect to the case of a higher derivative kinetic term for a Higgs particle in linearly realised EWSB: 
\begin{itemize}
\item[-] The number of  effective couplings modified is larger than in the linear case in Eqs.~\eqref{L_renorm_linear_fermions}-\eqref{VboxPhi}, a characteristic feature already explored previously in other settings~\cite{Brivio:2013pma}.
\item[-] Specifically, {\it couplings involving gauge particles {\bf are} now modified with respect to their SM values};  in addition to anomalous gauge-fermion interactions, particularly interesting anomalous Higgs couplings to two  (HVV) and three gauge bosons (HVVV), two Higgs-two gauge bosons (HHVV) and quartic gauge couplings (VVVV) are expected. The pure-gauge and gauge-Higgs anomalous couplings will be analyzed in detail in the next sections; they constitute a new tool to disentangle experimentally an elementary versus a dynamical nature of the Higgs particle, in the presence of higher-derivative kinetic terms. 
\end{itemize}

\subsection{Analysis via EOM}
The alternative method of applying directly  to the operator $\cP_{\square h}$ in the original non-linear Lagrangian Eq.~(\ref{Lchiral}) the standard field theory EOM for the $h$ field, 
\begin{equation}\label{EOM_h}
  \square h = -\dfrac{\delta V(h)}{\delta h}-\dfrac{v+h}{2}\tr[\VL_\mu\VL^\mu]-\dfrac{1}{\sqrt2}\left(\bar{Q}_L\UH\cY Q_R+\text{h.c.}\right),
\end{equation}
leads  to the same effective low-energy Lagrangian at first order on $c_{\square h}$ than that in Eqs.~\eqref{L_renorm_chiral_ferm}-\eqref{Vhchiral}, obtained above via the LW procedure, as it can be easily checked.  Again, the correction to the scalar potential requires to impose $c_{\square h}<0$ to ensure that the potential remains bounded from below.

\subsubsection*{Higgs potential}

\begin{figure}[h]\centering
 \includegraphics[width=0.8\textwidth]{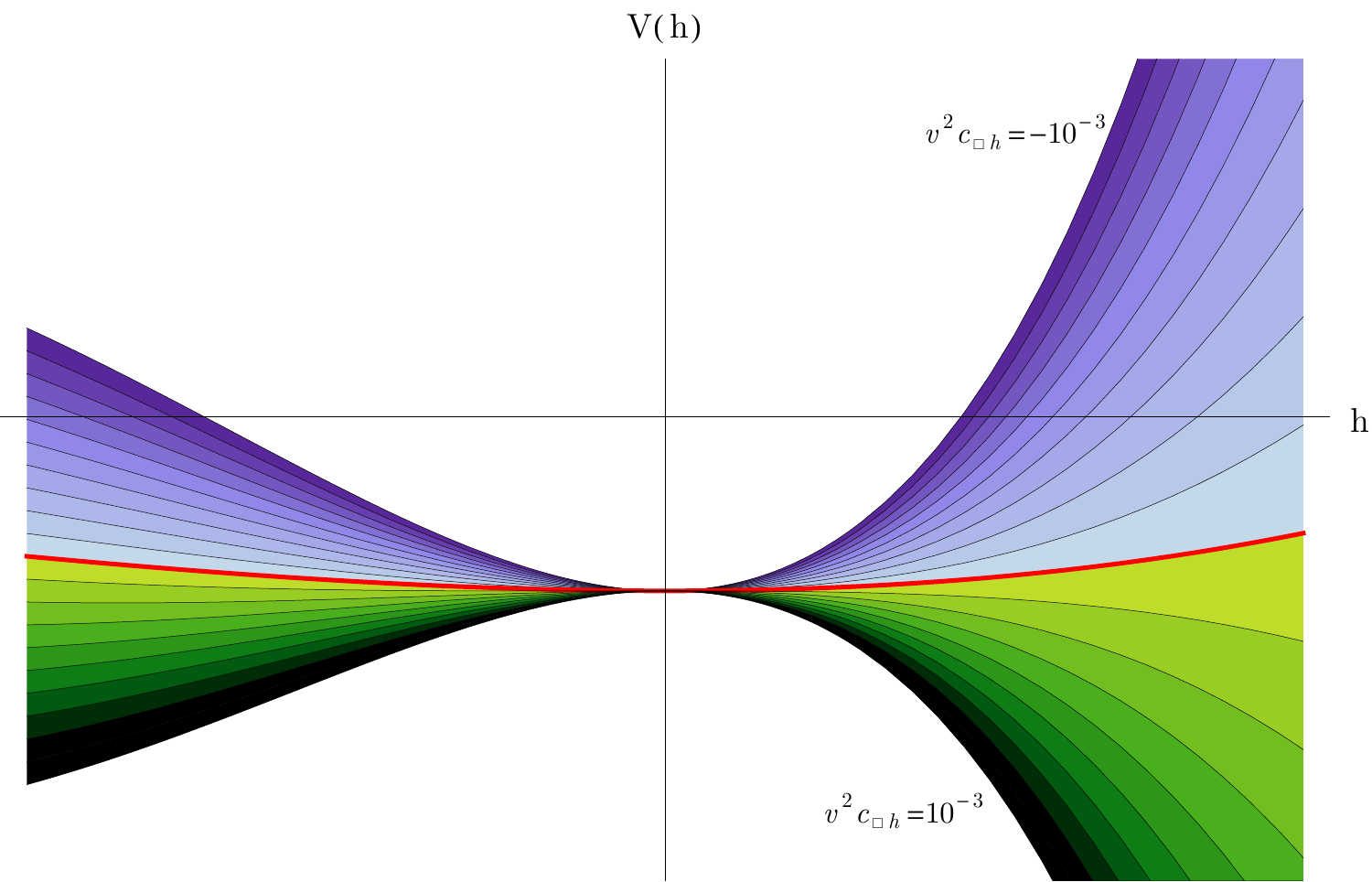}
 \caption{\it The scalar potential in the chiral Lagrangian for different values of the coefficient $v^2c_{\square h}$. The solid red line denotes the SM and the interline spacing is $\Delta(v^2 c_{\square\Phi})=7.5\cdot 10^{-5}$.}\label{fig.chiralV}
\end{figure}

Figure~\ref{fig.chiralV} shows the dependence of the shape of the scalar potential on the perturbative parameter $c_{\square h}$: for negative values the SM vacuum $\langle h \rangle=0$ is still a minimum, while for positive values the potential is not bounded from below and moreover the SM vacuum is turned into a maximum.

%
%
\newpage
\section{Chiral versus linear effective operators}

The linear operator  $\cO_{\square \Phi}$ involves gauge fields  in its structure  - see Eq.~(\ref{OboxPhi}),  contrary to the chiral effective operator $\cP_{\square h}$ defined in Eq.~(\ref{Oboxh}). 
Nevertheless, the addition of the former operator to the SM Lagrangian turned out to give {\it no contribution} to couplings involving gauge fields, while the chiral operator $\cP_{\square h}$ does. This seemingly paradoxical state of affairs and the consistency of the results can be ascertained by establishing the exact correspondence between both operators, which we find to be given by:
\begin{align}
\cO_{\square\Phi}=  &
\dfrac{1}{2}(\square h)^2
 +\dfrac{(v+h)^2}{8}\left(\tr[\VL_\mu\VL^\mu]\right)^2
 +\dfrac{v+h}{2}\tr[\VL_\mu\VL^\mu]\derp_\nu\derp^\nu h
 -\tr[\VL_\mu\VL_\nu]\derp^\mu h \derp^\nu h+\nn\\
 &-\dfrac{(v+h)^2}{4}\tr[(\DL_\mu\VL^\mu)^2]
 -(v+h)\tr[\VL_\nu\DL_\mu\VL^\mu]\derp^\nu h\label{Obh_siblings}\\
 =&\cP_{\square h} 
 + v^2\left(\dfrac{1}{8}\cP_6
 +\dfrac{1}{4}\cP_7
 -\cP_8
 -\dfrac{1}{4}\cP_9
 -\dfrac{1}{2}\cP_{10}\right)_{linear \,\mathcal{F}}\nn\,.
\end{align}
The right hand-side of Eq.~(\ref{Obh_siblings}) describes a combination of 
the non-linear operator  $\cP_{\square h}$ and a particular set of 
independent effective operators of the non-linear basis as determined in Ref.~\cite{Brivio:2013pma},
defined by  
\beq
\begin{aligned}
 \cP_6 &= \left(\tr[\VL_\mu\VL^\mu]\right)^2\,\mathcal{F}_6(h)\,, &
 \cP_7 &= \tr[\VL_\mu\VL^\mu]\,\square\mathcal{F}_7(h)\,,\\
 \cP_8 &= \tr[\VL_\mu\VL_\nu]\,\derp^\mu\mathcal{F}_8(h)\derp^\nu\mathcal{F'}_8(h)\,,&
 \cP_9 &= \tr[(\DL_\mu\VL^\mu)^2]\,\mathcal{F}_9(h)\,,\\
 \cP_{10} &= \tr[\VL_\nu\DL_\mu\VL^\mu]\,\derp^\nu\mathcal{F}_{10}(h)\,,
 \label{P6-10}
\end{aligned}
\eeq
where  the generic -model dependent- $\cF_i(h)$ functions are often parametrised as~\cite{Alonso:2012px,Brivio:2013pma} 
\begin{equation}
\cF_i(h) = 1+ 2 a_i\dfrac{h}{v}+b_i\dfrac{h^2}{v^2}\dots 
\label{def-F}
\end{equation}
 The subscript ``$linear\,\mathcal{F}$" in the right-hand side of Eq.~(\ref{Obh_siblings}) indicates that the equality holds when the arbitrary functions $\mathcal{F}_i(h)$ take the specific linear-like dependence -see Ref~\cite{Brivio:2013pma}~\footnote{In that reference, powers of the $\xi$ parameter -which refers to ratios of scales involved- were extracted from the definition of the operator coefficients; we will refrain here from doing so, and adopt the simple notation in Eq.~(\ref{deltaL}).\label{fotenote2}}-
\begin{equation}
\mathcal{F}_6(h)= \mathcal{F}_7(h)= \mathcal{F}_9(h)= \mathcal{F}_{10}(h)\overset{{linear\,\mathcal{F}}}{=}
(1+h/v)^2\,, \qquad  \mathcal{F}_8(h)= \mathcal{F'}_8(h)\overset{{linear\,\mathcal{F}}}{=} (1+h/v)\,.
\label{linearF}
\end{equation}
Strictly speaking, in a general chiral Lagrangian  the definition of $\cP_{\square h}$ should also contain  a  $\mathcal{F}_{\square h}(h)$ factor on the right hand side of Eq.~\eqref{Oboxh}~\cite{Buchalla:2013rka,Brivio:2013pma};  it would be superfluous 
to keep track of $\mathcal{F}_{\square h}(h)$ here, though, as we will restrain the analysis  to couplings involving at most two Higgs particles, which is tantamount to setting $\mathcal{F}_{\square h}(h)=1$ in the phenomenological analysis.

Taken separately, $\cP_{\square h}$ as well as each of the five operators in Eq.~(\ref{P6-10}) do induce deviations on the SM expectations for couplings involving gauge bosons. 
 Eq.~(\ref{Obh_siblings}) implies nevertheless that the gauge contributions of these six operators will exactly cancel in any physical observable when their relative weights are given by 
\beq
v^2c_{\square h} =
8 c_6 =
4 c_7 =
- c_8=
-4 c_9=
-2 c_{10}\,. 
\label{c-condition}
\eeq
We have explicitly checked such cancellations in several examples of physical transitions; 
 Appendix~\ref{ZZ} describes the particular case of $Z Z \to ZZ$ scattering, for illustration.

%

 \section{Signatures and Constraints}
 \label{Sect:SignCons}

Tables \ref{tab:F}, \ref{tab:V}, \ref{tab:VH}, and
\ref{tab:VHH} list all couplings involving up to four particles
that receive contributions from the effective linear operator
$\cO_{\square\Phi}$  or any of its chiral
siblings $\cP_{\square h}$ and $\cP_{6-10}$. We work at first order in the operator coefficients, which are left arbitrary in those tables; the $\mathcal{F}_i(h)$ functionals are also assumed generic as  defined in Eq.~(\ref{def-F}).
 For the sake of comparison,  a SM-like potential is taken for both the linear and chiral operators; the extension to a general scalar potential for the chiral expansion can be found in Appendix \ref{Generic_chiral_potential} and has no significant impact.

It turns out that $\cO_{\square\Phi}$ gives no tree-level contribution to couplings involving gauge particles as argued earlier, while instead $\cP_{\square h}$ and $\cP_{6-10}$ are shown to have a strong impact on a large number of gauge couplings. On the other side, anomalous four-fermion interactions are induced by both $\cO_{\square\Phi}$ and $\cP_{\square h}$, even if with distinct patterns. 


\begin{table}[h!]
\hspace{-0.5cm}
\renewcommand{\arraystretch}{2}
 \begin{tabular}{*5{|>{$}c<{$}}|}
 \hline
  \text{Fermionic couplings}& \text{Coeff.}& \text{SM value}&	\text{Chiral} &	\text{Linear: $\cO_{\square\Phi}$}\\\hline

   h\left(\bar{u}_L Y_U u_R+\bar{d}_L Y_D d_R+\text{h.c.}\right)&
   -\frac{1}{\sqrt2}&
   1&
   -m_h^2 c_{\square h}&
   -m_h^2 c_{\square \Phi}\\
    
   h^2\left(\bar{u}_L Y_U u_R+\bar{d}_L Y_D d_R+\text{h.c.}\right)&
   -\frac{1}{v\sqrt2}&
   -&
   -\frac{3m_h^2}{2}c_{\square h}&
   -\frac{3m_h^2}{2}c_{\square \Phi}\\\hline
   
    
    Z_\mu Z^\mu\left(\bar{u}_LY_U u_R+\bar{d}_LY_D d_R+\text{h.c.}\right)&
    -\frac{g^2v}{4\sqrt2\ct^2}&
    -&
     c_{\square h}&
    -
    \\
    
    W^+_\mu W^{-\mu}\left(\bar{u}_LY_U u_R+\bar{d}_LY_D d_R+\text{h.c.}\right)&
    -\frac{g^2v}{2\sqrt2}&
    -&
     c_{\square h}&
    -
    \\\hline


    \left(\bar{u}_L Y_U u_R\right)^2+\left(\bar{u}_RY^\dag_U u_L\right)^2&
    \frac{1}{4}&
    -&
     c_{\square h}&
    -
    \\    
    
    \left(\bar{u}_L Y_U u_R\right)\left(\bar{u}_RY^\dag_U u_L\right)&
    \frac{1}{2}&
    -&
     c_{\square h}&
    2c_{\square \Phi}
    \\

    \left(\bar{d}_L Y_D d_R\right)^2+\left(\bar{d}_RY^\dag_D d_L\right)^2&
    \frac{1}{4}&
    -&
    c_{\square h}&
    -
    \\
    
    \left(\bar{d}_L Y_D d_R\right)\left(\bar{d}_RY^\dag_D d_L\right)&
    \frac{1}{2}&
    -&
     c_{\square h}&
    2c_{\square \Phi}
    \\
   
   \left(\bar{u}_L Y_U u_R\right)\left(\bar{d}_RY^\dag_D d_L\right)+\text{h.c.}&
    \frac{1}{2}&
    -&
     c_{\square h}&
    -
    \\
    
    \left(\bar{u}_L Y_U u_R\right)\left(\bar{d}_LY_D d_R\right)+\text{h.c.}&
    \frac{1}{2}&
    -&
     c_{\square h}&
    2c_{\square \Phi}
    \\
    
   \left(\bar{u}_L Y_D d_R\right)\left(\bar{d}_LY_U u_R\right)+\text{h.c.}&
    -1&
    -&
    -&
    c_{\square \Phi}
    \\

    \left(\bar{u}_L Y_D d_R\right)\left(\bar{d}_RY_D^\dag u_L\right)+\left(\bar{d}_L Y_U u_R\right)\left(\bar{u}_RY_U^\dag d_L\right)&
    1&
    -&
    -&
    c_{\square \Phi}
    \\
    \hline
    \end{tabular}
\caption{\it Effective couplings involving fermions 
generated by  the linear operator $\cO_{\square\Phi}$ 
and its chiral siblings $\cP_{\square h}$ and $\cP_{6-10}$.
For illustration only the couplings involving quark pairs are listed, 
although similar interactions  involving lepton pairs are induced.}
\label{tab:F}
\end{table}


\boldmath
\subsection{Effects from $\cO_{\square\Phi}$}
\unboldmath

The only impact of  $\cO_{\square\Phi}$ on present Higgs and gauge boson observables is to generate 
the universal shift in the Higgs coupling to fermions shown in the first
line of Table \ref{tab:F}.  Equivalently, in the notation in Refs.~\cite{Azatov:2012bz,Espinosa:2012im,Plehn:2012iz,Aad:2013wqa,Chatrchyan:2013lba}, in which the deviations of the Yukawa couplings and the gauge kinetic terms from SM predictions were parametrised as
\begin{align}
\mathcal{L}_{Yukawa}&\equiv-\frac{v}{\sqrt2}\left(\bar{Q}_L\UH \cY Q_R+\text{h.c.}\right) \left(1+c\frac{h}{v}+\dots\right)\,,
\label{LYukawa}\\
\mathcal{L}_{gauge-kinetic}&\equiv-\frac{v^2}{4}\tr(\VL_\mu \VL^\mu) \left(1+2 a \frac{h}{v}+ b\frac{h^2}{v^2}+\dots\right)\,,\label{LGauge}
\end{align}
the shift induced by the operator $\cO_{\square\Phi}$ reads 
\beq 
c\equiv\kappa_f\equiv 1+\Delta_f= 1-m_h^2 c_{\square
  \Phi}\,.
\label{eq:aclin}
\eeq
while
\beq
a\equiv\kappa_V\equiv 1+\Delta_V =1\,,\qquad\qquad\qquad b=1\,. 
\eeq
In Ref.~\cite{Corbett:2012ja,Brivio:2013pma}, a general analysis of the constraints on
departures of the Higgs couplings strength from SM expectations used all available collider and EW precision data, and it
was found that
\begin{equation}
-0.55 \leq \Delta_f \leq 0.25 \: ,
\label{eq:hboundB}
\end{equation}
at 90\% CL after marginalizing over all other effective couplings.
Eq.~(\ref{eq:hboundB}) constrains $ m_h^2 c_{\square \Phi}$, in
addition to any combination of coefficients of other dimension--six
operators which may also modify universally the Higgs couplings to
fermions, see for instance Ref~\cite{Brivio:2013pma}. 

When only $\cO_{\square \Phi}$ is added to the SM Lagrangian, Eq.~(\ref{eq:hboundB}) translates into the bound $c_{\square \Phi}\lesssim1.6\cdot10^{-5} \GeV^{-2}$.  This constraint is quantitatively quite weak, a fact due to present sensitivity. For illustration,  it could be rephrased as a lower limit of $~250 \GeV$ on the Higgs doublet LW partner mass. It shows that the bound obtained is of the order of magnitude of the constraints established by previous analyses, which considered direct production in colliders and/or indirect contributions to EW precision data and flavour data
~\cite{Rizzo:2007ae,Rizzo:2007nf,Dulaney:2007dx,Alvarez:2008za,Alvarez:2008ks,Underwood:2008cr,Carone:2008bs,Carone:2009nu}, setting a lower bound for the LW scalar partner mass of $445\GeV$.


\boldmath
\subsection{Effects from $\cP_{\square h}$ and $\cP_{6-10}$}
\unboldmath

Tables \ref{tab:V}, \ref{tab:VH} and \ref{tab:VHH} illustrate that 
$\cP_{\square h}$ generates tree-level corrections to the gauge
boson self-couplings, as well as to gauge-Higgs couplings. Note that some of these interactions would not be induced by {\it any} $d=6$ operator of a linear expansion, an example being the $ZZZZ$ interactions in Table~\ref{tab:V}; other signals absent in both the SM and $d=6$ linear expansions, and thus unique to the leading order chiral expansion,  can be found in Appendix \ref{ChiralVsLinearAPP}. They constitute a strong tool to disentangle  a strong underlying EW dynamics from a linear one.

 The effects stemming from the operators $\cP_{6-10}$, which are also siblings of the linear operator
$\cO_{\square\Phi}$, are displayed in these tables for gauge two-point functions (VV), triple gauge vertices (TGV) and VVVV couplings. As previously discussed, the tree-level contributions to physical amplitudes induced by that set of chiral operators cancel if
the conditions in Eqs.~(\ref{linearF}) and (\ref{c-condition})  are satisfied.
Notwithstanding, for generic
values of the coefficients of $\cP_{\square h}$ and $\cP_{6-10}$, 
some  signatures  characteristic of a non-linearly realised electroweak symmetry breaking are expected, as those 
discussed next.

\begin{table}[h!]
\centering
\renewcommand{\arraystretch}{2}
 \begin{tabular}{*5{|>{$}c<{$}}|}
 \hline
  \text{VV, TGV and VVVV}& \text{Coeff.}& \text{SM value}&	\text{Chiral} &	\text{Linear: $\cO_{\square\Phi}$} \\\hline
  (\derp_\mu Z^{\mu}) (\derp_\nu Z^{\nu}) &
    -\frac{g^2}{2\ct^2}&
    -&
    c_9& 
    -
    \\
  
  (\derp_\mu W^{+\mu}) (\derp_\nu W^{-\nu})&
    -g^2&
    -&
    c_9& 
    -
    \\
    \hline

   i (\derp_\mu W^{-\mu})(Z_\nu W^{+\nu})+\text{h.c.}&
   \frac{e^2 g}{\ct^2}&
  -&
  c_9&
   -\\
 
  i (\derp_\mu W^{-\mu})(A_\nu W^{+\nu})+\text{h.c.}&
  -e g^2&
  -&
  c_9&
 -\\
 \hline
    \left(Z_\mu Z^\mu\right)^2& 
    \frac{g^4}{32\ct^4}&
    -&
     v^2c_{\square h}+8c_6&
    -
    \\

    \left(W^+_\mu W^{-\mu}\right)^2& 
    -\frac{g^2}{2}&
    1&
    -m_W^2 c_{\square h}-2g^2 c_6&
    -
    \\

    \left(W^+_\mu W^{-\mu}\right)\left(Z_\nu Z^\nu\right)& 
    -g^2\ct^2&
    1&
    -\frac{m_Z^2}{2} c_{\square h}-\frac{g^2}{\ct^4} c_6&
    -
    \\
    
    \left(W^+_\mu Z^{\mu}\right)\left(W^-_\nu Z^\nu\right)& 
    g^2\ct^2&
    1&
    -\frac{e^2 \st^2}{\ct^4}c_9&
    -
    \\
    
    \left(W^+_\mu A^{\mu}\right)\left(W^-_\nu A^\nu\right)& 
    e^2g^2&
    1&
    -c_9&
    -
    \\
    
    \left(W^+_\mu A^{\mu}\right)\left(W^-_\nu Z^\nu\right)+\text{h.c.}& 
    eg\ct&
    1&
    \frac{e^2}{\ct^2}c_9&
    -
    \\
    
    \hline
\end{tabular}
\caption{\it Anomalous pure-gauge couplings involving two, three and four gauge bosons, 
induced by  the chiral operators $\cP_{\square h}$ and $\cP_{6-10}$,  in contrast with the non-impact of their linear sibling 
$\cO_{\square\Phi}$.}
\label{tab:V}
\end{table}

\begin{table}[h!]
\centering
\renewcommand{\arraystretch}{2}
 \begin{tabular}{*5{|>{$}c<{$}}|}
 \hline
  \text{HVV and HVVV}& \text{Coeff.}& \text{SM value}&	\text{Chiral} &	\text{Linear: $\cO_{\square\Phi}$}\\\hline

    Z_\mu Z^\mu h& 
    \frac{vg^2}{4\ct^2}&
    1&
    -m_h^2 c_{\square h} &
    -
    \\
  
   Z_\mu Z^\mu \square h& 
    -\frac{g^2}{2\ct^2}&
    -&
    \frac{2c_7a_7}{v} &
    -
    \\  
  
    (\derp_\mu Z^\mu) (\derp_\nu Z^\nu) h&
    -\frac{g^2}{2\ct^2}&
    -&
    \frac{2c_9a_9}{v}& 
    -
    \\ 
    
    (\derp_\mu Z^\mu)(Z_\nu \derp^\nu h)&
    -\frac{g^2}{2\ct^2}&
    -&
    \frac{2c_{10}a_{10}}{v}&
    -\\
  
    W^+_\mu W^{-\mu} h& 
    \frac{vg^2}{2}&
    1&
    -m_h^2  c_{\square h}&
    -
    \\

    W^+_\mu W^{-\mu} \square h& 
    -g^2&
    -&
    \frac{2c_7a_7}{v}&
    -\\

   (\derp_\mu W^{+\mu}) (\derp_\nu W^{-\nu}) h&
    -g^2&
    -&
    \frac{2c_9a_9}{v}& 
    -
    \\

    (\derp_\mu W^{+\mu})(W^-_\nu \derp^\nu h)+\text{h.c.}&
    -\frac{g^2}{2}&
    -&
    \frac{2c_{10}a_{10}}{v}&
    -\\
    
\hline  

i (\derp_\mu W^{-\mu})(Z_\nu W^{+\nu})h+\text{h.c.}&
  \frac{e^2 g}{\ct^2}&
  -&
  \frac{2c_9a_9}{v}&
 -\\
 
  i (\derp_\mu W^{-\mu})(A_\nu W^{+\nu})h+\text{h.c.}&
  -e g^2&
  -&
  \frac{2c_9a_9}{v}&
 -\\

  i (Z_\mu W^{+\mu})(W^-_\nu \derp^\nu h)+\text{h.c.}&
  -\frac{e^2g}{2\ct}&
  -&
  \frac{2c_{10}a_{10}}{v}&
  -\\

   i (A_\mu W^{+\mu})(W^-_\nu \derp^\nu h)+\text{h.c}&
   \frac{eg^2}{2}&
   -&
   \frac{2c_{10}a_{10}}{v}&
   -\\

\hline
\end{tabular}
\caption{\it 
Anomalous effective couplings of the Higgs particle to two or three gauge bosons, 
induced by  the chiral operators $\cP_{\square h}$ and $\cP_{6-10}$,  in contrast with the non-impact of their linear sibling 
$\cO_{\square\Phi}$.}
\label{tab:VH}
\end{table}

\begin{table}[h!]
\centering
\renewcommand{\arraystretch}{2}
 \begin{tabular}{*5{|>{$}c<{$}}|}
 \hline
  \text{ H$^2$VV couplings}& \text{Coeff.}& \text{SM value}&	\text{Chiral}&	\text{Linear: $\cO_{\square\Phi}$}\\\hline


     Z_\mu Z^\mu h^2& 
    \frac{g^2}{8\ct^2}&
    1&
    -5m_h^2 c_{\square h}&
    -
    \\

   Z_\mu Z^\mu \square (h^2)& 
    -\frac{g^2}{2\ct^2}&
    -&
    \frac{c_7b_7}{v^2} &
    -
    \\  

    Z_\mu Z_\nu \derp^\mu h \derp^\nu h& 
    -\frac{g^2}{2\ct^2}&
    -&
   \frac{4c_8a_8a'_8}{v^2}&
    -
    \\
    
    (\derp_\mu Z^\mu) (\derp_\nu Z^\nu) h^2&
    -\frac{g^2}{2\ct^2}&
    -&
    \frac{c_9b_9}{v^2}& 
    -
    \\ 
    
    (\derp_\mu Z^\mu)(Z_\nu \derp^\nu h)h&
    -\frac{g^2}{2\ct^2}&
    -&
    \frac{2c_{10}b_{10}}{v^2}&
    -\\
    
     W^+_\mu W^{-\mu} h^2& 
    \frac{g^2}{4}&
    1&
    -5m_h^2 c_{\square h} &
    -\\

    W^+_\mu W^{-\mu} \square (h^2)& 
    -g^2&
    -&
    \frac{c_7b_7}{v^2}&
    -\\
   
   W^+_\mu W^-_\nu \derp^\mu h \derp^\nu h& 
    -g^2&
    -&
    \frac{4c_8a_8a'_8}{v^2}&
    -\\
    
    (\derp_\mu W^{+\mu}) (\derp_\nu W^{-\nu}) h^2&
    -g^2&
    -&
    \frac{c_9b_9}{v^2}& 
    -
    \\

   (\derp_\mu W^{+\mu})(W^-_\nu \derp^\nu h)h+\text{h.c.}&
    -\frac{g^2}{2}&
    -&
    \frac{2c_{10}b_{10}}{v^2}&
    -\\
    
    \hline
\end{tabular}
\caption{\it 
Anomalous effective couplings involving two Higgs particles and two gauge bosons, 
induced by  the chiral operators $\cP_{\square h}$ and $\cP_{6-10}$,  in contrast with the non-impact of their linear sibling 
$\cO_{\square\Phi}$.}
\label{tab:VHH}
\end{table}

From Tables \ref{tab:VH} and \ref{tab:F} it follows that $\cP_{\square h}$
yields a universal correction to the SM Higgs couplings to gauge bosons
and fermions. Furthermore, in present Higgs data the Higgs state is
on-shell and, in this case, $\cP_{7}$ gives also a correction to the
SM-like HVV couplings, while the modifications generated by
$\cP_{9}$ and $\cP_{10}$ vanish for on-shell $W$ and $Z$ gauge bosons or
massless fermions. Thus these corrections can be cast as, in the notation of Eqs.~(\ref{LYukawa}) and (\ref{LGauge}),
\beq 
a\equiv\kappa_V\equiv1+\Delta_V=
1-\frac{m_h^2}{v^2}(v^2c_{\square h}+4 c_7 a_7)\,,\qquad
c\equiv\kappa_f\equiv1+\Delta_f= 1-m_h^2 c_{\square h}\,,
\label{eq:acchiral}
\eeq
with $b_7=0.$
The general constraints resulting from present Higgs and other data~\cite{Corbett:2012ja,Brivio:2013pma} apply as well here. For instance, 
if the coefficients of operators contributing only to the
SM-like HVV coupling -- such as $c_7a_7$ above -- cancel, the bound on $\Delta_V$ and $\Delta_f$  
becomes, at 90\% CL,
\begin{equation}
-0.33 \leq \Delta_f=\Delta_V \leq 0.33 \,,
\end{equation}
which translates into a bound $c_{\square h}\lesssim 2.1\cdot 10^{-5} \GeV^{-2}$.

\subsubsection*{Off-shell Higgs mediated gauge boson pair production}

\begin{figure}
  \centering
  \includegraphics[width=0.45\textwidth]{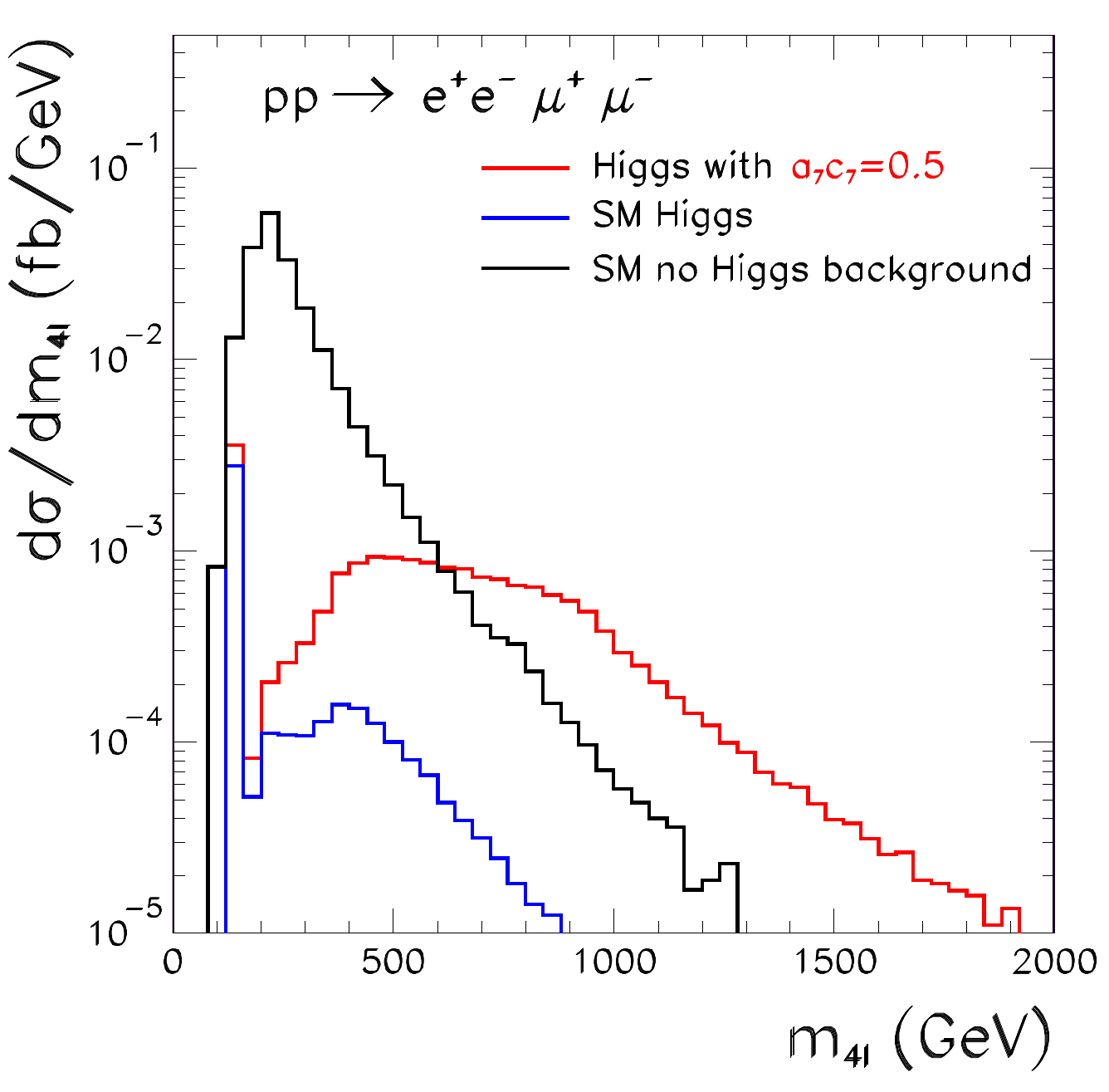}
  \includegraphics[width=0.45\textwidth]{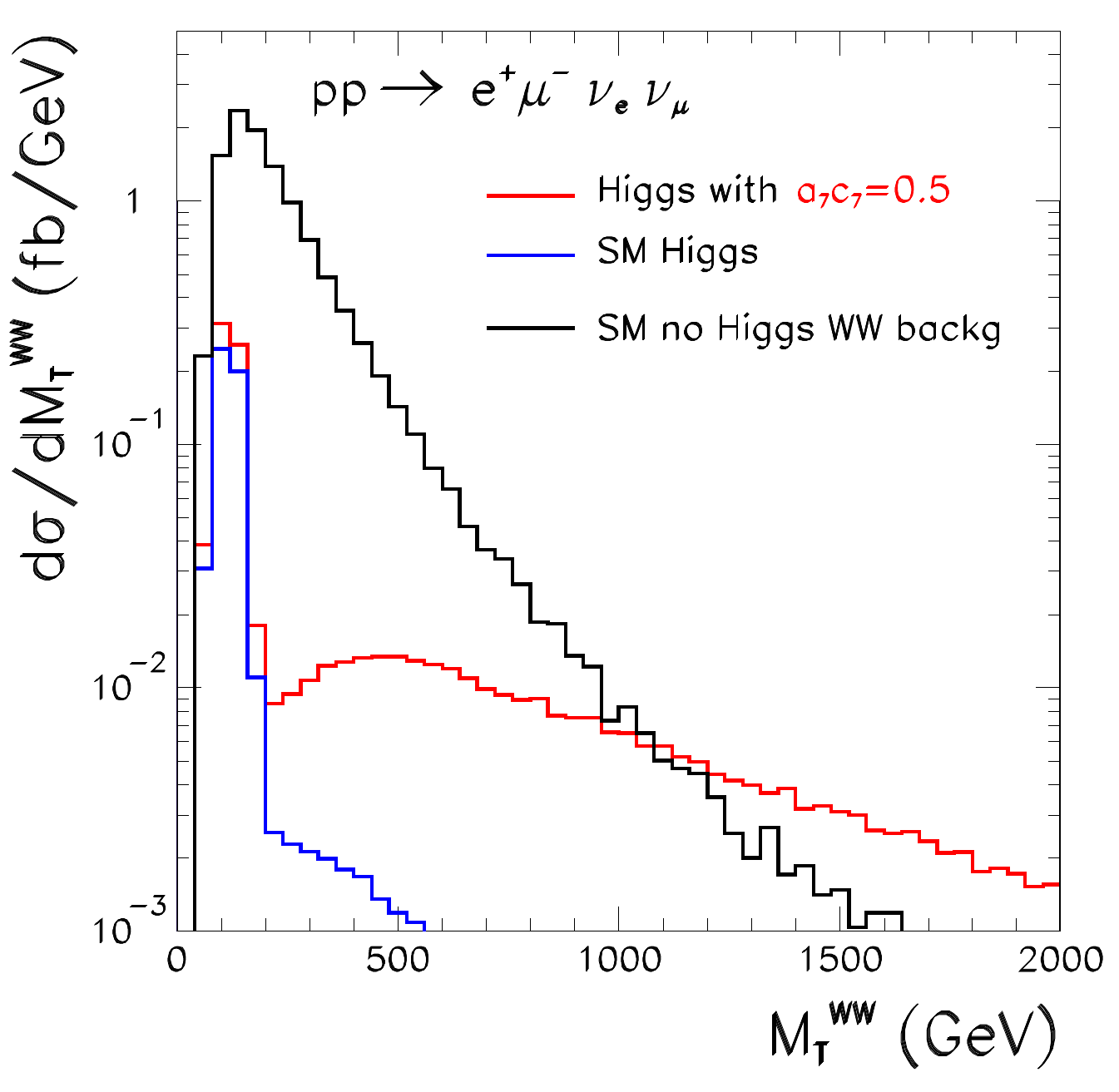}
  \caption{\it The left panel presents the four lepton invariant mass
    spectrum for the process $p p \to e^+ e^- \mu^+ \mu^-$. The right
    panel contains the $WW$ transverse mass distribution of the
    process $p p \to e^+ \nu_e \mu^- \nu_\mu$.  In both panels the
    black line stands for the SM leading-order contribution while the
    blue (red) one represents the SM ($\cP_7$) higher-order
    contribution given by Eq.  (\ref{eq:ggvv}).  In this figure we
    assumed a center--of--mass energy of 13 TeV and $c_7a_7 = 0.5$.  }
\label{fig:zzww}
\end{figure}

Potentially more interesting, $\cP_{7}$ leads to a new  contribution to the
production of electroweak gauge--boson pairs $ZZ$ and $W^+ W^-$
through
\begin{equation}
   g g \to h^\star \to ZZ \hbox{ or } W^+ W^- \;,
\label{eq:ggvv}
\end{equation}
where the Higgs boson is off--shell~\cite{Gainer:2014hha,Englert:2014aca}. 
For the sake of illustration, we consider the $ZZ$ pair production
with one $Z$ decaying into $e^+ e^-$ while the other into $\mu^+
\mu^-$. The left panel of Figure~\ref{fig:zzww} depicts the leading-order SM contribution to
\[
   p p \to e^+ e^- \mu^+ \mu^-\,,
\]
together with the SM higher-order  and  $\cP_7$ contributions through the $ZZ$ channel  
in Eq.\ (\ref{eq:ggvv}). The results presented in this figure were
obtained  assuming a center-of-mass energy at the LHC of $13$ TeV, and requiring that all leptons have transverse momenta in excess
of 10 GeV, that they are central ($|\eta|<2.5$) and that the same-flavour opposite-charge lepton pairs reconstruct the $Z$ mass
($|M_{\ell^+ \ell^-} - M_Z|< 5 $ GeV).  In presenting the $\cP_7$
effects a coupling $c_7 a_7 = 0.5$ was assumed, which is compatible with the
presently available Higgs data.
Also, since the goal here is to illustrate the effects of $\cP_7$, we did not take into account 
the SM higher-order contribution to $g g \to
e^+ e^- \mu^+ \mu^-$ which interferes with the off-shell Higgs one; for
further details see Ref.~\cite{Campbell:2013una} and references
therein.

The results in the left panel of the Figure~\ref{fig:zzww} show that $\cP_7$ leads to an enhancement of the
off-shell Higgs cross section with respect to the SM expectations at
high four-lepton invariant masses. In fact, the scattering amplitude
grows so fast that at some point unitarity is
violated~\cite{Gainer:2014hha}, and the introduction of some unitarization 
procedure will tend to diminish the excess.  Nevertheless, even without an 
unitarization procedure,  the expected number of events above the 
leading order SM background induced by  $\cP_7$ is shown to be very small, meaning that unraveling the $\cP_7$ contribution 
will be challenging.

We have analyzed as well the process
\[
   p p \to e^+ \nu_e \mu^- \nu_\mu\,,
\]
that can proceed via the $W^+W^-$ channel in Eq.~(\ref{eq:ggvv}). In
the right panel of Figure~\ref{fig:zzww}  the corresponding cross section is depicted as
a function of the $WW$ transverse mass
\begin{equation}
  M_T^{WW} 
= \biggl[ \left( \sqrt{(p_T^{\ell^+\ell^{\prime -}})^2 + 
m^2_{\ell^+ \ell^{\prime -}}}   
              + \sqrt{\slashed{p}_T^2 + m^2_{\ell^+\ell^{\prime -} }} \right)^2
- (\vec{p}_T^{~\ell^+\ell^{\prime -}} + 
\vec{\slashed{p}_T}  )^2 \biggr]^{1/2} \,,
\label{eq:mtww}
\end{equation}
where $\vec{\slashed{p}_T}$ stands for the missing transverse momentum
vector, $\vec{p}_T^{~\ell^+\ell^{\prime -}}$ is the transverse
momentum of the pair $\ell^+ \ell^{\prime -}$ and
$m_{\ell^+\ell^{\prime -}}$ is the $\ell^+ \ell^{\prime -}$ invariant
mass. Here $\ell=e$ or $\mu$. 
The transverse momentum and rapidity cuts used were the same than those for the left panel.
As expected, an enhancement of the $g g \to e^+ \nu_e \mu^-
\nu_\mu$ cross section is induced by the operator $\cP_7$.  
Analogously to the case of $ZZ$ production, the SM
leading-order contribution dominates but for large $M_T^{WW}$; the
expected signals from the excess due to $\cP_7$ will  be thus very difficult
to observe.

\subsubsection*{Corrections to four gauge boson scattering}

As can be seen in  Tables \ref{tab:V} and \ref{tab:VH} 
the combination $v^2c_{\square h}+8c_6$ 
generates the anomalous quartic vertex $ZZZZ$ that is not present
in the SM. Moreover,  the same combination gives anomalous contributions 
to the $ZZW^+W^-$  and $W^+W^-W^+W^-$. These are genuinely four gauge
boson effects which do not induce any modification to triple gauge
boson couplings and, therefore, these coefficients are much less
constrained at present.

Nowadays the most stringent bounds on the coefficients of these
operators are indirect, from their one--loop contribution to the
electroweak precision data~\cite{Brunstein:1996fz}, in particular to
$\alpha\Delta T$ which at 90\% CL imply 
\begin{equation}
-0.23 \leq\frac{1}{8}
v^2 c_{\square h} +c_6 \leq 0.26\,.
\end{equation}
At the LHC with 13-14 TeV center-of-mass energy, they 
can be detected or constrained by combining their impact on the VBF channels
\begin{equation}
  p p \to jj W^+ W^-   \;\;\;\hbox{and}\;\;\;
    p p \to jj (W^+ W^+ + W^-W^-) \,,
\end{equation}
where $j$ stands for a tagging jet and the final state $W$'s decay
into electron or muon plus neutrino~\cite{Eboli:2006wa};  
the attainable 99\% CL limits on these couplings are
\begin{equation}
-1.2 \cdot 10^{-2} \leq \frac{1}{8}v^2 c_{\square h} +c_6<10^{-2} \, .
\end{equation}
Disregarding the contribution from $c_6$, this would translate into $c_{\square h}\lesssim1.3\cdot10^{-6} \GeV^{-2}$, which would suggest a sensitivity to the mass of the LW partner for the singlet Higgs in the chiral EWSB realization up to $\sim887 \GeV$.

Strictly speaking,  the relevant four gauge boson cross-section  also receives modifications induced by those operators
which correct the HVV and TGV vertices when the Higgs boson or a gauge boson is exchanged in the $s$, $t$ 
 or $u$ channels.
In principle, these ``triple vertex''  effects can be discriminated from the 
purely VVVV effects by their different  dependence  on the scattering angle 
of the final  state  gauge bosons. In practice, a detailed simulation will 
be required to establish the final sensitivity to all relevant
coefficients.

%
%

\section{Conclusions}

An effective coupling for bosons which is tantamount to a quartic kinetic energy  is a  full-rights member of the tower of  leading effective operators accounting for BSM physics in a model-independent way. This is so in both the linear and non-linear realizations of electroweak symmetry breaking, or in other words irrespective of whether the light Higgs particle corresponds to an elementary or a composite (dynamical) Higgs.  The corresponding higher derivative kinetic couplings, denoted here $\cO_{\square \Phi}$ and $\cP_{\square h}$, respectively, Eqs.~(\ref{OboxPhi}) and (\ref{Oboxh}),  are customarily not considered  but traded by others (e.g. fermionic ones) instead of being kept as  independent elements of a given basis. 

It is most pertinent to analyze those couplings directly, though, as they are related to intriguing and potentially very important solutions to ultraviolet issues, such as  the electroweak gauge hierarchy problem. The field theory challenges they rise constitute as well a fascinating theoretical conundrum. Their theoretical impact is ``diluted" and hard to track, though, when they are traded by combinations of other operators. 
On top of which, the present LHC data offer  increasingly rich and precise constraints on gauge and gauge-Higgs couplings, up to the point of becoming competitive with fermionic bounds in constraining BSM theories; this trend  may be further strengthened with the post-LHC facilities presently under discussion.

We have analyzed and compared in this paper $\cO_{\square \Phi}$ and $\cP_{\square h}$,   
 unravelling theoretical and experimental distinctive features.

On the theoretical side, two analyses have been carried in parallel and  compared: i) the Lee-Wick procedure of trading the second pole in the propagator by a ``ghost" scalar partner; ii)  the application of the EOM to the operator, trading it by other effective operators and resulting in an analysis which only requires standard field-theory tools. Both paths have been shown to be consistent, producing the same effective Lagrangian at leading order in the operator coefficient dependence. 

A most interesting property is that the physical impact differs for linearly versus non-linear EWSB realizations: departures from SM values for  quartic-gauge boson, Higgs-gauge boson and  fermion-gauge boson couplings are expected  only for the case of a dynamical Higgs, i.e. only from $\cP_{\square h}$ while not from $\cO_{\square \Phi}$; in addition, they induce a different pattern of deviations on Yukawa-like fermionic couplings and on the Higgs potential.  

Note that these distinctive signals of a dynamical origin of the Higgs particle would be altogether  missed if a  $d=6$ linear effective Lagrangian was used to evaluate the possible impact of an underlying strong dynamics, showing that in general a linear approach is not an appropriate tool to the task. Indeed,  for completeness we  identified all TGV, HVV and VVVV experimental signals which are unique in resulting from the leading chiral expansion, while they cannot be induced neither by SM couplings at tree-level {\it nor} by $d=6$ operators of the linear expansion: the TGV couplings $\Delta g^{\gamma}_{6}$, $\Delta g^{Z}_{5}$ and $\Delta g^{Z}_{6}$,  the HVV couplings  $\Delta g^{(4)}_{HVV}$, $\Delta g^{(5)}_{HVV}$ and $\Delta g^{(6)}_{HVV}$ and the VVVV couplings $\Delta g^{(1)}_{ZZ}$ and $\Delta g^{(5)}_{\gamma Z}$, 
with the quartic kinetic energy coupling for  non-linear EWSB scenarios $\cP_{\square h}$ contributing only to $\Delta g^{(1)}_{ZZ}$ among the above.  
The experimental search of that ensemble of couplings and their correlations (see Tables \ref{tabtgv}, \ref{tabhvv} and \ref{tab4V} in App.~\ref{ChiralVsLinearAPP}), constitute a superb window into chiral dynamics  associated to the Higgs particle.

To tackle  the origin of the different physical impact of  quartic derivative Higgs kinetic terms depending on the type of EWSB, 
we have explored and established the precise relation between the two couplings: it was shown that $\cO_{\square \Phi}$   corresponds to a specific combination of $\cP_{\square h}$ with five other non-linear operators.  

On the phenomenological analysis, the impact of $\cO_{\square \Phi}$, $\cP_{\square h}$ and $\cP_{6-10}$ has been scrutinised. All LHC Higgs and other data presently available  were used to constrain  the $\cO_{\square \Phi}$ and $\cP_{\square h}$ coupling strengths. Moreover, the impact of future 14 TeV LHC data on $pp\to\text{4 leptons}$ has been explored;
the operators under scrutiny  intervene in the process via  off-shell Higgs mediation in gluon-gluon fusion,  $g g \to h^\star \to ZZ \hbox{ or } W^+ W^-$, inducing excesses at high four-lepton invariant masses via the $ZZ$ channel, and at high values of the  $WW$ invariant mass in the $WW$ channel. The corrections expected at LHC through their impact on four gauge boson scattering, extracted combining information from vector boson fusion channels, $  p p \to jj W^+ W^-$  and
    $p p \to jj (W^+ W^+ + W^-W^-)$,  has been also discussed.  The  possibility that LHC may shed light on Lee-Wick theories through the type of analysis and signals discussed here is a fascinating perspective.

%
%

\section*{Acknowledgements}
We thank especially J. Gonzalez-Fraile for early discussions about  the
presence of new off-shell Higgs effects. 
We acknowledge partial support of the European Union network FP7 ITN
INVISIBLES (Marie Curie Actions, PITN-GA-2011-289442), of CiCYT
through the project FPA2009-09017, of CAM through the project HEPHACOS
P-ESP-00346, of the European Union FP7 ITN UNILHC (Marie Curie
Actions, PITN-GA-2009-237920), of MICINN through the grant
BES-2010-037869, of the Spanish MINECO Centro de Excelencia Severo
Ochoa Programme under grant SEV-2012-0249, and of the Italian
Ministero dell'Uni\-ver\-si\-t\`a e della Ricerca Scientifica through
the COFIN program (PRIN 2008) and the contract MRTN-CT-2006-035505.
The work of I.B. is supported by an ESR contract of the European Union
network FP7 ITN INVISIBLES mentioned above. The work of L.M. is
supported by the Juan de la Cierva programme (JCI-2011-09244). The
work of O.J.P.E. is supported in part by Conselho Nacional de
Desenvolvimento Cient\'{\i}fico e Tecnol\'ogico (CNPq) and by
Funda\c{c}\~ao de Amparo \`a Pesquisa do Estado de S\~ao Paulo
(FAPESP). The work of M.C.G-G is supported by USA-NSF grant PHY-09-6739, 
by CUR Generalitat de Catalunya grant
2009SGR502, by MICINN FPA2010-20807 and by
consolider-ingenio 2010 program CSD-2008-0037.

\newpage
  
\appendix \small

%
%
\boldmath
\section{Analysis with a generic chiral potential $V(h)$}
\unboldmath
\label{Generic_chiral_potential}

In the analysis performed in this paper the effective operators $\cP_{\square h}$ and $\cO_{\square\Phi}$ are assumed to be the only departures from the Standard Model present in the  chiral and linear Lagrangians, respectively.
However, the choice of a SM-like scalar potential might not appear satisfactory for the chiral case: \textit{a priori} $V(h)$ is a completely generic polynomial in the singlet field $h$, and the current lack of direct measurements of the triple and quartic self-couplings of the Higgs boson leaves room for a less constrained parametrization.

Therefore, it can be interesting to test the stability of our results against deviations of the scalar potential from the SM pattern.
To do this, we apply the Lee-Wick method to the Lagrangian in Eq.~(\ref{Lchiral}) although with the SM-like potential in Eq.~(\ref{VSM})  replaced by a generic one, 
\begin{equation}\label{chiral_potential}
 V(h)=  a_1h+\frac{m_h^2}{2}a_2 h^2+\frac{m_h^2}{2v}a_3 h^3 +\frac{m_h^2}{8v^2}a_4 h^4\,,
\end{equation}
where we choose to omit higher $h$-dependent terms, as the analysis remains at tree level and limited to interactions involving at most two Higgs particles.
The correction factor $a_2$ can always be reabsorbed in the definition of $m_h$, and will thus be fixed from the start  to $$a_2=1\,.$$
The comparison with the case described in Section~\ref{Composite_section} is straightforward  choosing, in addition, $a_3=a_4=1$ and $a_1=0$.  
The resulting mass-diagonal Lagrangian containing the LW field $\chi$ is: 
\begin{equation}
 \mathcal{L}^\chi = (\text{kin. terms})+ \mathcal{L}^\chi_Y+ \mathcal{L}^\chi_\text{gauge}-V(h,\chi)\,,
\end{equation}
with
\begin{align}
\mathcal{L}^\chi_Y &= -\frac{1}{\sqrt 2}(\bar{Q}_L \UH\cY Q_R+\text{h.c.})\left[1+ (1+2x)(h-\chi)\right]\,,\\
\mathcal{L}^\chi_\text{gauge} &=-\frac{1}{4}\tr[\VL_\mu\VL^\mu]\left[v^2+2v(1+2x)(h-\chi)+(1+4x)(h-\chi)^2\right]\,,\\
V(h,\chi) &=a_1 (1+2x) (h-\chi)
+\frac{m_h^2}{2}(1+2x)h^2
+\frac{m_h^2}{2}\left(1-\frac{1}{2x}+2x\right)\chi^2\nn\\
&\qquad
+\frac{m_h^2}{2v}a_3(1+6x)(h-\chi)^3
+\frac{m_h^2}{8v^2}a_4(1+8x)(h-\chi)^4\,,
\end{align}
where $x=-c_{\square h}m_h^2/2>0$.

Upon integrating out the heavy LW ghost, the following renormalized Lagrangian results:
\begin{equation}\label{L_renorm_chiral_fermions}
 \begin{aligned}
 \mathcal{L}_{\square h} &=\dfrac{1}{2}\derp_\mu h\derp^\mu h-\dfrac{1}{4}\ZZd\ZZu-\dfrac{1}{2}\WWd^+ W^{-\mu\nu}+i\bar{Q}\slashed{D}Q&+\mathcal{L}_{\square h}^\text{fer.}+\mathcal{L}_{\square h}^\text{gauge}- V_{\square h}(h)\,,
\end{aligned}
\end{equation}
where 
\begin{align}
 \mathcal{L}_{\square h}^\text{fer.}=&
 -\dfrac{1}{\sqrt2}\left(\bar{Q}_L\UH \cY Q_R+\text{h.c.}\right)
 \Bigg[v+ \left(1+2x\right)h
 +3 a_3 x \frac{h^2}{v}+a_4x \frac{h^3}{v^2}\Bigg]+\label{L_genV_f}\\
 &- \dfrac{x}{2m^2}\left(\bar{Q}_L\UH\cY Q_R+\text{h.c.}\right)^2-\frac{ x}{m_h^2}\dfrac{v+h}{\sqrt2 }\tr[\VL_\mu\VL^\mu] \left(\bar{Q}_L\UH\cY Q_R+\text{h.c.}\right)\,,\nn\\[5mm]
 \mathcal{L}_{\square h}^\text{gauge}=&
 -\dfrac{1}{4}\tr[\VL_\mu\VL^\mu]
 \Bigg[
(v+h)^2\left(1+4x\frac{h}{v}+2x h^2\right)+2x(v+h)\frac{h^2}{v^2}\left(3v(a_3-1)+h(a_4-1)\right)\Bigg]+\nn\\
 & -\frac{x}{4m_h^2}\left(v+h\right)^2\tr[\VL_\mu\VL^\mu]^2\,,\\[5mm]
 V_{\square h}(h)=&\dfrac{m_h^2}{2}h^2
 +a_1(1+2x)h
 +\frac{m_h^2}{2v}\left[a_3(1+4x)+\frac{2a_1 x}{m_h^2 v}(a_4+a_3 -3a_3^2)\right]h^3\label{L_genV_V}\\
 & +\frac{m_h^2}{8v^2}\left[a_4(1+6x)+2x\left(9a_3^2+\frac{a_1a_4}{m^2 v}(2-3a_3)\right)\right]h^4
 +\frac{3a_3a_4m_h^2x}{2v^3}h^5+\frac{a_4^2m_h^2x}{4v^4}h^6\,.\nn
\end{align}


\subsection*{Phenomenological impact}
 Assuming that the departures from unity of the $a_i$ parameters are small (of order $c_{\square h}$ at most), we can replace
\begin{equation}
\begin{aligned}
 	 a_1&\to \Delta a_1\,,\qquad\qquad&		 	 a_i&\to 1+\Delta a_i\,,\qquad i=3,4
\end{aligned}
\end{equation}
and expand the renormalized Lagrangian~\eqref{L_renorm_chiral_fermions} up to first order in $x$ \emph{and} in the $\Delta_i$'s.
Restricting for practical reasons to  vertices with up to four legs, the list of couplings that are modified is very reduced and only includes terms in the scalar potential:
\beq
\begin{aligned}
&-\dfrac{m_h^2}{2v}(1+4x+\Delta a_3)h^3\,,\\
&-\dfrac{m_h^2}{8v^2}(1+24x+\Delta a_4)	h^4\,,\\
&-\Delta a_1h\,.\\
\end{aligned}
\eeq
In consequence, upon the assumption that possible departures of the scalar potential from a SM-like form are  quantitatively at most of the same order as $c_{\square h}$, those contributions would not affect the numerical analysis presented in the text. 

%
%
\boldmath
\section{Impact of $\cO_{\square\Phi}$ versus  $\cP_{\square h}$ on $ZZ\to ZZ$ scattering }
\label{ZZ}
\unboldmath

This Appendix provides an illustrative example of how the contributions of the chiral operators $\cP_{\square h}\,, \cP_{6-10}$ to physical amplitudes combine to reproduce those of the linear operator $\cO_{\square\Phi}$, once the conditions~\eqref{c-condition} and~\eqref{linearF} are imposed.

Let us consider the elastic scattering of  two $Z$ gauge bosons. This process is not affected by $\cO_{\square\Phi}$, therefore the corrections induced by the six chiral operators are expected to cancel exactly, upon assuming~\eqref{c-condition} and~\eqref{linearF}.

Assuming the external $Z$ bosons are on-shell, the only Feynman diagrams containing deviations from the Standard Model are the following
\begin{figure}[h!]\centering

\begin{fmffile}{ZZscatter}
\begin{align}\label{A_h_mediated}
\mathcal{A}_s+\mathcal{A}_t+\mathcal{A}_u=&\quad
\parbox{3cm}{\centering
\begin{fmfgraph*}(60,40)
\fmfleft{i1,i2}
\fmfright{o2,o1}
  \fmf{boson}{i2,v1}
  \fmf{boson}{i1,v1}
  \fmf{dashes,label=$h$}{v1,v2}
  \fmf{boson}{v2,o1}
  \fmf{boson}{v2,o2}
\fmfv{lab=$Z_1$,l.angle=180}{i2}
\fmfv{lab=$Z_2$,l.angle=180}{i1}
\fmfv{lab=$Z_3$,l.angle=0}{o1}
\fmfv{lab=$Z_4$,l.angle=0}{o2}
\end{fmfgraph*}}+
\parbox{3cm}{\centering
\begin{fmfgraph*}(60,40)
\fmfleft{i1,i2}
\fmfright{o2,o1}
  \fmf{boson}{i2,v1}
  \fmf{boson}{i1,v2}
  \fmf{dashes,label=$h$}{v1,v2}
  \fmf{boson}{v1,o1}
  \fmf{boson}{v2,o2}
\fmfv{lab=$Z_1$,l.angle=180}{i2}
\fmfv{lab=$Z_2$,l.angle=180}{i1}
\fmfv{lab=$Z_3$,l.angle=0}{o1}
\fmfv{lab=$Z_4$,l.angle=0}{o2}
\end{fmfgraph*}}+
\parbox{3cm}{\centering
\begin{fmfgraph*}(60,40)
\fmfleft{i1,i2}
\fmfright{o2,o1}
  \fmf{boson}{i2,v1}
  \fmf{boson}{i1,v2}
  \fmf{dashes,label=$h$}{v1,v2}
  \fmf{boson}{v1,o1}
  \fmf{boson}{v2,o2}
\fmfv{lab=$Z_1$,l.angle=180}{i2}
\fmfv{lab=$Z_2$,l.angle=180}{i1}
\fmfv{lab=$Z_3$,l.angle=0}{o2}
\fmfv{lab=$Z_4$,l.angle=0}{o1}
\end{fmfgraph*}}
\\[.7cm]
\mathcal{A}_{4Z} =&\quad
\parbox{3cm}{\centering
\begin{fmfgraph*}(60,40)
\fmfleft{i1,i2}
\fmfright{o2,o1}
  \fmf{boson}{i2,v1}
  \fmf{boson}{i1,v1}
  \fmf{boson}{v1,o1}
  \fmf{boson}{v1,o2}
\fmfv{lab=$Z_4$,l.angle=0}{o1}
\fmfv{lab=$Z_3$,l.angle=0}{o2}
\fmfv{lab=$Z_1$,l.angle=180}{i1}
\fmfv{lab=$Z_2$,l.angle=180}{i2}
\end{fmfgraph*}}\label{A_4Z}
\end{align}

\end{fmffile}
\end{figure}

For the amplitudes depicted in~\eqref{A_h_mediated}, the relevant couplings are $ZZh$ and $ZZ\square h$ (see Table~\ref{tab:VH}), and the contributions from each channel turn out to be
\begin{align}
\mathcal{A}_s &=
-(\e_1\cdot\e_2)(\e^*_3\cdot\e^*_4)\dfrac{i}{s-m_h^2}\dfrac{4m_Z^4}{v^2}\left(1-2m_h^2c_{\square h}+\dfrac{8s}{v^2}c_7a_7\right)\,,\\
\mathcal{A}_t &=
-(\e_1\cdot\e^*_3)(\e_2\cdot\e^*_4)\dfrac{i}{t-m_h^2}\dfrac{4m_Z^4}{v^2}\left(1-2m_h^2c_{\square h}+\dfrac{8s}{v^2}c_7a_7\right)\,,\\
\mathcal{A}_u &=
-(\e_1\cdot\e^*_4)(\e^*_3\cdot\e_2)\dfrac{i}{u-m_h^2}\dfrac{4m_Z^4}{v^2}\left(1-2m_h^2c_{\square h}+\dfrac{8s}{v^2}c_7a_7\right)\,,
\end{align}

where $\e_1, \e_2$ denote the polarizations of the incoming $Z$ bosons, and $\e^*_3, \e^*_4$ those of the outgoing ones. 

Imposing the constraints $c_7=v^2c_{\square h}/4$, from eq.~\eqref{c-condition} and $a_7=1$ from eq.~\eqref{linearF}, the dependence on the exchanged momentum drops from the non-standard part of the amplitudes:
\beq
\begin{aligned}
\mathcal{A}_h= \mathcal{A}_s+\mathcal{A}_t+\mathcal{A}_u=&\,
-\dfrac{4im_Z^4}{v^2}\left[
\dfrac{(\e_1\cdot\e_2)(\e^*_3\cdot\e^*_4)}{s-m_h^2}+
\dfrac{(\e_1\cdot\e^*_3)(\e_2\cdot\e^*_4)}{t-m_h^2}+
\dfrac{(\e_1\cdot\e^*_4)(\e_2\cdot\e^*_3)}{u-m_h^2}\right]+\\
&-\dfrac{8im_Z^4}{v^2}c_{\square h}
\Big[(\e_1\cdot\e_2)(\e^*_3\cdot\e^*_4)+(\e_1\cdot\e^*_3)(\e_2\cdot\e^*_4)+(\e_1\cdot\e_4)(\e_2\cdot\e^*_3)\Big]\,.
\end{aligned}
\eeq

The diagram~\eqref{A_4Z} contains only the four-point vertex $ZZZZ$ (see table~\ref{tab:V}), and gives
\beq
\begin{aligned}
\mathcal{A}_{4Z}=&\dfrac{32im_Z^4}{v^4}\left(c_6+\dfrac{v^2}{8}c_{\square h}\right) \Big[(\e_1\cdot\e_2)(\e^*_3\cdot\e^*_4)+(\e_1\cdot\e^*_3)(\e_2\cdot\e^*_4)+(\e_1\cdot\e_4)(\e_2\cdot\e^*_3)\Big]=\\
=&
\dfrac{8im_Z^4}{v^2}c_{\square h} \Big[(\e_1\cdot\e_2)(\e^*_3\cdot\e^*_4)+(\e_1\cdot\e^*_3)(\e_2\cdot\e^*_4)+(\e_1\cdot\e_4)(\e_2\cdot\e^*_3)\Big]\,.
\end{aligned}
\eeq
In the second line the condition~\eqref{c-condition} has been assumed, which imposes $v^2c_{\square h}=8c_6$. 

The neat correction to the Standard Model amplitude for $ZZ$ scattering induced by the chiral operators $\cP_{\square h}\,,\cP_{6-10}$ is finally proved to vanish, as
\begin{equation}
 \Delta\mathcal{A} = \Delta\mathcal{A}_h + \mathcal{A}_{4Z} 
 =0\,.
\end{equation}

%
%

\newpage
\section{Chiral versus linear couplings}
\label{ChiralVsLinearAPP}

In this appendix, we gather the departures from SM couplings in TGV, HVV and VVVV vertices, which are expected from the leading order tower of chiral scalar and/or gauge operators (which includes $\cP_{\square h}$ and $\cP_{6-10}$ discussed in this manuscript), as well as from any possible chiral or  $d=6$ linear coupling  which may affect those same vertices at leading order of the respective effective expansions. 
Their comparison allows a straightforward  identification of which signals may point to a strong dynamics underlying EWSB, being free from SM or $d=6$ linear operators contamination. In Tables \ref{tabtgv}, \ref{tabhvv} and \ref{tab4V} below:
\begin{itemize}
\item[-]  The $\cO_{\square\Phi}$ , $\cP_{\square h}$ and $\cP_{6-10}$ operators are defined as in Eqs.~(\ref{OboxPhi}), (\ref{Oboxh})and (\ref{P6-10}),  while for all other couplings mentioned -linear or chiral-  the  naming follows that  in Ref.~\cite{Brivio:2013pma}, to which we refer the reader. 
\item[-] {\it All} operator coefficients appearing in the tables below are defined as in Eq.~(\ref{deltaL}).  In comparison with the definitions in Ref.~\cite{Brivio:2013pma} this means that:  i) the coefficient of the chiral operator $\cP_{\square h}$  has been rescaled, see footnotes \ref{fotenote1} and \ref{fotenote2}; ii) the  $d=6$ linear operator coefficients $f_i$ in Ref.~\cite{Corbett:2012ja,Brivio:2013pma} are related to those in the tables below as follows:
\begin{equation}
c_i=f_i/\Lambda^2\,.
\end{equation}
\end{itemize}
As discussed in the text, new anomalous vertices related to a quartic kinetic energy for the Higgs particle include as well HHVV couplings and new corrections to fermionic vertices. We leave for a future publication the corresponding comparison between the complete linear and chiral bases. When referring below to the SM, only tree-level contributions are considered.

\subsection{TGV couplings}

The CP-even sector of the Lagrangian that
describes TGV couplings can be parametrized as
\begin{align}
\LL_{WWV} =& - \,i g_{WWV} \Bigg\{ 
g_1^V \Big( W^+_{\mu\nu} W^{- \, \mu} V^{\nu} - 
W^+_{\mu} V_{\nu} W^{- \, \mu\nu} \Big) 
   \,+\, \kappa_V W_\mu^+ W_\nu^- V^{\mu\nu}\, \label{eqclassical} \\
& -  ig_5^V \epsilon^{\mu\nu\rho\sigma}
\left(W_\mu^+\partial_\rho W^-_\nu-W_\nu^-\partial_\rho W^+_\mu\right)
V_\s \,+\,
 g_{6}^V \left(\derp_\mu W^{+\mu} W^{-\nu}-\derp_\mu W^{-\mu} W^{+\nu}\right)
V_\nu  \Bigg\}\,,\nn
\end{align}
where $V \equiv \{\gamma, Z\}$ and $g_{WW\gamma} \equiv e=g
\sin\theta_W$, $g_{WWZ} = g \cos\theta_W$. The SM values for the phenomenological parameters defined in this expression are $g_{1}^{Z,\gamma}=\kappa_{Z,\gamma}=1$ and $g_5^{Z,\gamma}=g_{6}^{Z,\gamma}=0$.
The resulting TGV corrections are gathered in Table \ref{tabtgv}. For instance, while $\Delta g_{6}^\gamma$ and $\Delta g_{6}^Z$ cannot be induced  by any linear $d=6$ operators, they receive contributions from the operators $\cP_{6-10}$ discussed in this manuscript. Barring fine-tunings and one-loop effects, a detection of such couplings with sizeable strength would point to a non-linear realization of EWSB.

\begin{table}[h!]
\centering
\footnotesize
\begin{tabular}{|c||c||c||c|}
\hline
&&\multicolumn{1}{c}{}&\\[-2mm]
& Coeff. & Chiral & Linear \\[2mm]
& $\times e^2/\st^2$& 
& $\times v^2$ \\[2mm]
\hline
\hline
&&&\\[-3mm]
$\Delta\kappa_\gamma$ 
&$1$
&$-2c_1+2c_2+c_3-4c_{12}+2c_{13}$
&$\frac{1}{8}(c_W+c_B- 2 c_{BW})$
\\[2mm]
$\Delta g_{6}^\gamma$
&$1$
&$-c_{9}$
&$-$
\\[2mm]
\hline
&&&\\[-3mm]
$\Delta g_1^Z$
&$\frac{1}{\ct^2}$
&$\frac{\sdt^2}{4e^2\cdt}c_T +\frac{2\st^2}{\cdt}c_1+ c_3$
&
$ 
\frac{1}{8}c_W + \frac{\st^2}{4\cdt}c_{BW}-\frac{\sdt^2}{16e^2\cdt} c_{\Phi,1} 
$
\\[2mm]
$\Delta\kappa_Z$
&$1$
&$\frac{\st^2}{e^2\cdt}c_T+\frac{4\st^2}{\cdt}c_1
-\frac{2\st^2}{ct^2}c_2+ c_3-4c_{12}+2c_{13}$
&$\frac{1}{8}c_W -\frac{\st^2}{8 ct^2}c_B 
+\frac{\st^2}{2\cdt}c_{BW} -\frac{\st^2}{4e^2\cdt}c_{\Phi,1}$
\\[2mm]
$\Delta g_5^Z$
&$\frac{1}{\ct^2}$
&$c_{14}$
&$-$
\\[2mm]
$\Delta g_{6}^Z$
&$\frac{1}{\ct^2}$
&$\st^2c_{9}-c_{16}$
&$-$
\\[2mm]
\hline
\end{tabular}
\caption{\it Effective couplings parametrizing the $V W^+ W^-$
vertices defined in Eq.~(\ref{eqclassical}).  The coefficients in the
second column are common to both the chiral and linear expansions. The
 third  column lists the specific contributions from the
operators in the chiral basis.
For comparison, the last column exhibits the corresponding
contributions from linear $d=6$ operators.}
\label{tabtgv}
\end{table}

\subsection{HVV couplings}

The Higgs to two gauge bosons couplings can be phenomenologically parametrized as 
\begin{align}
\LL_\text{HVV}\equiv&
\phantom{+}g_{Hgg} \,G^a_{\mu\nu} G^{a\mu\nu} h+
g_{H \gamma \gamma} \, A_{\mu \nu} A^{\mu \nu} h+ 
g^{(1)}_{H Z \gamma} \, A_{\mu \nu} Z^{\mu} \partial^{\nu} h + 
g^{(2)}_{H Z \gamma} \, A_{\mu \nu} Z^{\mu \nu} h \nn
\\
&+g^{(1)}_{H Z Z}  \, Z_{\mu \nu} Z^{\mu} \partial^{\nu} h + 
g^{(2)}_{H Z Z}  \, Z_{\mu \nu} Z^{\mu \nu} h+
g^{(3)}_{H Z Z}  \, Z_\mu Z^\mu h + g^{(4)}_{H Z Z}  \, Z_\mu Z^\mu \Box h\nn
\\
&+ g^{(5)}_{H Z Z}  \, \derp_\mu Z^\mu Z_\nu \derp^\nu h+ g^{(6)}_{H Z Z}  \, \derp_\mu Z^\mu \derp_\nu Z^\nu  h\label{eqlhvv}
\\
&+ g^{(1)}_{H W W}  \, \left (W^+_{\mu \nu} W^{- \, \mu} \partial^{\nu} h + \hc \right) +
g^{(2)}_{H W W}  \, W^+_{\mu \nu} W^{- \, \mu \nu} h+
g^{(3)}_{H W W}  \, W^+_{\mu} W^{- \, \mu} h\nn
\\
&+g^{(4)}_{H W W}  \, W^+_\mu W^{-\mu} \Box h+ g^{(5)}_{H W W}  \, 
\left(\derp_\mu W^{+\mu} W^-_\nu \derp^\nu h+\hc\right)+ g^{(6)}_{H W W}  \, \derp_\mu W^{+\mu} \derp_\nu W^{-\nu}  h\,,\nn
\end{align}
where $V_{\mu \nu} = \partial_\mu V_\nu - \partial_\nu V_\mu$
with $V=\{A, Z, W, G\}$.  
Separating the contributions into SM ones plus corrections, $g_i^{(j)}\simeq  g_i^{(j)SM} + \Delta g_i^{(j)}\,,$
it turns out that
\beq
\begin{aligned}
 g^{(3)SM}_{HZZ}=\frac{m_Z^2}{v}\,\,,\qquad\qquad
g^{(3)SM}_{HWW}=\frac{2m_Z^2\ct^2}{v}\,, 
\end{aligned}
\eeq
while the tree-level SM value for all other couplings in
Eq.~(\ref{eqlhvv}) vanishes. 

While  $\cP_{\square h}$ may induce a departure from SM expectations in two HVV couplings, $\Delta g^{(3)}_{H Z Z}$ and $\Delta g^{(3)}_{H W W}$, Table \ref{tabhvv} shows that those signals could be mimicked by some $d=6$ linear operators. 
On the contrary, a putative detection of $\Delta g^{(4)}_{H VV}$ couplings may arise from the $\cP_7$ operator discussed in this manuscript while neither from the SM not  any linear $d=6$ couplings, and would thus be a smoking gun for a non-linear nature of EWSB realization; the same applies to $\Delta g^{(5)}_{H VV}$ from  
$\cP_{10}$, and to $\Delta g^{(6)}_{H VV}$ from  
$\cP_{9}$. 

\begin{table}[ht!]
\hspace*{-2cm}
\footnotesize
\begin{tabular}{|c||c||c||c|}
\hline
&&\multicolumn{1}{c}{}&\\[-2mm] 
& Coeff. & Chiral & Linear \\[2mm]
& $\times e^2/4v$& & $\times v^2$ \\[2mm]
\hline
\hline
&&&\\[-3mm]
$\Delta g_{Hgg}$ & $\frac{g^2_s}{e^2}$ & $-2c_Ga_G$  &$-4c_{GG}$ \\[2mm]
$\Delta g_{H \gamma \gamma}$ & $1$  & $-2(c_Ba_B+c_Wa_W)+ 8c_1a_1+8 c_{12}a_{12}$ &$ -(c_{BB}+c_{WW})+c_{BW}$  \\[2mm]
$\Delta g^{(1)}_{H Z \gamma}$ & $\frac{1}{\sdt}$  & $-8 (c_5a_5+2 c_4a_4)-16 c_{17}a_{17}$  & $2(c_W-c_B) $\\[2mm]
$\Delta g^{(2)}_{H Z \gamma}$ & $\frac{\ct}{\st}$  & $4\frac{\st^2}{\ct^2}c_Ba_B-4c_Wa_W+8\frac{\cdt}{\ct^2}c_1a_1+16 c_{12}a_{12}$ & $2\frac{\st^2}{\ct^2} c_{BB} -2c_{WW}+\frac{\cdt}{\ct^2} c_{BW}$  \\[2mm]
$\Delta g^{(1)}_{H Z Z}$ & $\frac{1}{\ct^2}$  & $-4\frac{\ct^2}{\st^2}c_5a_5+8 c_4a_4-8 \frac{\ct^2}{\st^2} c_{17}a_{17}$  & $ \frac{\ct^2}{\st^2} c_W+c_B$ \\[2mm]
$\Delta g^{(2)}_{H Z Z}$ & $-\frac{\ct^2}{\st^2}$ &$2\frac{\st^4}{\ct^4} c_Ba_B+2 c_Wa_W+8\frac{\st^2}{\ct^2} c_1a_1-8c_{12}a_{12}$ & $\frac{\st^4}{\ct^4}c_{BB}+c_{WW}+\frac{\st^2}{\ct^2} c_{BW}$ \\[2mm]
$\Delta g^{(3)}_{H Z Z}$ & $\frac{m_Z^2}{e^2}$ & $-2c_H+2c_C(2a_C-1)-8c_T(a_T-1)
-4m_h^2c_{\square h}
$  &$c_{\Phi,1}+ 2 c_{\Phi,4}-2 c_{\Phi,2}$ \\[2mm]
$\Delta g^{(4)}_{H Z Z}$ & $-\frac{1}{\sdt^2}$ & $16 c_7a_{7}+32 c_{25}a_{25}$&  $-$\\[2mm] 
$\Delta g^{(5)}_{H Z Z}$ & $-\frac{1}{\sdt^2}$ & $16c_{10}a_{10}+32c_{19}a_{19}$ 
&$-$\\[2mm] 
$\Delta g^{(6)}_{H Z Z}$ & $-\frac{1}{\sdt^2}$ & $16c_9a_{9}+32c_{15}a_{15}$ &$-$\\[2mm] 
$\Delta g^{(1)}_{H WW}$ & $\frac{1}{\st^2}$  & $-4 c_5a_5$   & $c_W$ \\[2mm]
$\Delta g^{(2)}_{H WW}$ & $\frac{1}{\st^2}$ & $-4 c_Wa_W$   & $-2c_{WW}$ \\[2mm]
$\Delta g^{(3)}_{H WW}$ & $\frac{m_Z^2\ct^2}{e^2}$ & $-4c_H+4c_C(2a_C-1)+\frac{32e^2}{\cdt} c_1
+\frac{16\ct^2}{\cdt}c_T
-8m_h^2c_{\square h}-\frac{32e^2}{\st^2}\,c_{12}$  & 
$\frac{-2(3\ct^2-\st^2)}{\cdt}c_{\Phi,1}+ 4 c_{\Phi,4}-4 c_{\Phi,2}+\frac{4  e^2}{\cdt}  c_{BW}$\\[2mm]
$\Delta g^{(4)}_{H W W}$ & $-\frac{1}{\st^2}$ & $8c_7a_{7}$  &
$-$\\[2mm] 
$\Delta g^{(5)}_{H W W}$ & $-\frac{1}{\st^2}$ & $4c_{10}a_{10}$ & $-$ \\[2mm] 
$\Delta g^{(6)}_{H W W}$ & $-\frac{1}{\st^2}$ & $8c_9a_{9}$  &$-$\\[2mm] 
\hline
\end{tabular}
\caption{\it Higgs-gauge bosons couplings as defined in
Eq.~(\ref{eqlhvv}). The coefficients in the second column are common
to both the chiral and linear expansions.The
 third  column lists the specific contributions from the
operators in the chiral basis.
For comparison, the last column exhibits the corresponding
contributions from linear $d=6$ operators.}
\label{tabhvv}
\end{table}

\subsection{VVVV couplings}
 
The effective Lagrangian for VVVV couplings reads
\beq
\begin{split}
\LL_{4X}\, \equiv \, g^2 & \Bigg\{ \, g^{(1)}_{ZZ} (Z_\mu Z^\mu)^2 
\,+\, g^{(1)}_{WW}\, W^+_\mu W^{+\mu} W^-_\nu W^{-\nu} \,-\, 
g^{(2)}_{WW} (W^+_\mu W^{-\mu})^2 \,  \\
& \,+ \, 
g^{(3)}_{VV'} W^{+\mu} W^{-\nu}\left( V_\mu V'_\nu + V'_\mu V_\nu \right) 
\,-\, 
g^{(4)}_{VV'} W^+_\nu W^{-\nu} V^\mu V'_\mu \, \\   
& \,+\, 
ig^{(5)}_{VV'} e^{\mu\nu\rho\s} W^+_\mu W^-_\nu V_\rho V'_\s \Bigg\}\,,
\end{split}
\label{eq4v}
\eeq
where $VV'=\{\gamma\gamma, \gamma Z, ZZ\}$. At tree-level in the SM, the following couplings are non-vanishing:
\beq
\begin{aligned}
g^{(1)SM}_{WW}&=\frac{1}{2}\,,\qquad
&g^{(2)SM}_{WW}&=\frac{1}{2}\,,\qquad
&g^{(3)SM}_{ZZ}&=\frac{\ct^2}{2}\,,\qquad
&g^{(3)SM}_{\g\g}&=\frac{\st^2}{2}\,,\\
g^{(3)SM}_{Z\g}&=\frac{\sdt}{2}\,,\qquad
&g^{(4)SM}_{ZZ}&=\ct^2\,,\qquad
&g^{(4)SM}_{\g\g}&=\st^2\,,\qquad
&g^{(4)SM}_{Z\g}&=\sdt\,,
\end{aligned}
\eeq

Table \ref{tab4V} shows the impact on the couplings in Eq.~(\ref{eq4v}) of the leading non-linear versus linear operators.  While  $\cP_{\square h}$ and  $\cP_{6}$ 
may induce  $\Delta g^{(2)}_{WW}$ and $\Delta g^{(4)}_{ZZ}$ couplings, the table shows that those signals could be mimicked by some $d=6$ linear operators. 
On the contrary, the 4Z coupling $\Delta g^{(1)}_{ZZ}$ is induced by $\cP_{\square h}$, while it vanishes in the SM and in any linear $d=6$ expansion. A detection of $\Delta g^{(1)}_{ZZ}$ would thus be a beautiful smoking gun of a non-linear nature of EWSB realization, which may simultaneously indicate a quartic kinetic energy for the Higgs scalar of LW theories (although $\Delta g^{(1)}_{ZZ}$ may also be induced by other chiral operators, including $\cP_6$ as discussed towards the end of Sect.~\ref{Sect:SignCons}).

\begin{table}[ht!]
\hspace*{-9mm}
\footnotesize
\centering
\renewcommand{\arraystretch}{2}
\begin{tabular}{|*2{>{$}c<{$}||}>{$}c<{$}||>{$}c<{$}|}
\hline
& \text{Coeff. }& \text{Chiral}  &  \text{Linear} \\[-1mm]
&\times e^2/4\st^2&  & \times v^2 \\
\hline
\hline
\Delta g^{(1)}_{WW}&
1&  \frac{\sdt^2}{e^2\cdt}c_T+\frac{8\st^2}{\cdt}c_1 +4 c_3 +
 2c_{11}-16c_{12} +8 c_{13}
 &\frac{c_W}{2}+\frac{\st^2}{\cdt}c_{BW}-\frac{\sdt^2}{4\cdt e^2}c_{\Phi1}
\\
\Delta g^{(2)}_{WW}&
1&  \frac{\sdt^2}{e^2\cdt}c_T+\frac{8\st^2}{\cdt}c_1 +4 c_3 -4 c_{6}
-\frac{v^2}{2}c_{\square h} -2c_{11} -16 c_{12} +8 c_{13}
&\frac{c_W}{2} +\frac{\st^2}{\cdt}c_{BW}-\frac{\sdt^2}{4\cdt e^2}c_{\Phi1}\\\hline 

\Delta g^{(1)}_{ZZ}&
\frac{1}{\ct^4}& c_{6}+\frac{v^2}{8}c_{\square h} +  c_{11} +2 c_{23} +2 c_{24} 
+ 4 c_{26}  & -\\
\Delta g^{(3)}_{ZZ}&
\frac{1}{\ct^2} &  \frac{\sdt^2\ct^2}{e^2\cdt}c_T
+\frac{2\sdt^2}{\cdt}c_1 +4\ct^2 c_3-2\st^4 c_{9}+
2c_{11}+4\st^2c_{16} + 2 c_{24}
& \frac{c_W\ct^2}{2} +\frac{\sdt^2}{4\cdt}c_{BW}-\frac{\sdt^2\ct^2}{4e^2\cdt}c_{\Phi1}\\ %
\Delta g^{(4)}_{ZZ}&
\frac{1}{\ct^2}& \frac{2\sdt^2\ct^2}{e^2\cdt}c_T
+\frac{4\sdt^2}{\cdt}c_1 +8\ct^2 c_3- 4c_{6}
-\frac{v^2}{2}c_{\square h} - 4c_{23} 
& c_W\ct^2+2\frac{\sdt^2}{4\cdt}c_{BW}-\frac{\sdt^2\ct^2}{2e^2\cdt}c_{\Phi1}\\\hline %
\Delta g^{(3)}_{\g\g}&
\st^2 & - 2c_{9} 
& -\\ \hline %
\Delta g^{(3)}_{\g Z}& 
\frac{\st}{\ct}& \frac{\sdt^2}{e^2\cdt}c_T+\frac{8\st^2}{ \cdt}c_1 + 4c_3 + 4\st^2 c_{9}
  - 4c_{16} & \frac{c_W}{2} +\frac{\st^2}{\cdt}c_{BW}-\frac{\sdt^2}{4\cdt e^2}c_{\Phi1} \\%
\Delta g^{(4)}_{\g Z}&
\frac{\st}{\ct} &  \frac{2\sdt^2}{e^2\cdt}c_T
+\frac{16\st^2}{ \cdt}c_1 +8 c_3  & c_W+2\frac{\st^2}{\cdt}c_{BW}-\frac{\sdt^2}{2\cdt e^2}c_{\Phi1} \\
\Delta g^{(5)}_{\g Z}&
\frac{\st}{\ct}&   8c_{14}   & - \\\hline
 \end{tabular}
\caption{\it Effective couplings parametrizing the vertices of four gauge 
bosons defined in Eq.~(\ref{eq4v}). The
 third  column lists the specific contributions from the
operators in the chiral basis.
For comparison, the last column exhibits the corresponding
contributions from linear $d=6$ operators.
}
\label{tab4V}
\end{table}

Summarising this appendix, some experimental signals are unique in resulting from the leading chiral expansion, while they cannot be induced neither by the SM at tree-level {\it nor} by $d=6$ operators of the linear expansion; among those analyzed here they are
\begin{itemize}
\item[-] the TGV couplings $\Delta g^{\gamma}_{6}$, $\Delta g^{Z}_{5}$, and $\Delta g^{Z}_{6}$, 
\item[-] the HVV couplings  $\Delta g^{(4)}_{HVV}$, $\Delta g^{(5)}_{HVV}$, and $\Delta g^{(6)}_{HVV}$,
\item[-] the VVVV couplings $\Delta g^{(1)}_{ZZ}$, and $\Delta g^{(5)}_{\gamma Z}$, 
\end{itemize}
with the quartic kinetic energy coupling for  non-linear EWSB scenarios $\cP_{\square h}$ contributing only to $\Delta g^{(1)}_{ZZ}$ among the above. $\Delta g^{(3)}_{\gamma \gamma}$ does not receive contributions from $d=6$ linear operators, but it is induced by three-level SM effects. The experimental search of that ensemble of couplings, with the correlations among them following from Tables \ref{tabtgv}, \ref{tabhvv} and \ref{tab4V}, constitute a fascinating window into chiral dynamics  associated to the Higgs particle.

\normalsize
\providecommand{\href}[2]{#2}\begingroup\raggedright\endgroup

\end{document}